\documentclass[sigconf]{acmart}

\AtBeginDocument{%
  }

\copyrightyear{2025}
\acmYear{2025}
\setcopyright{cc}
\setcctype{by}
\acmConference[FAccT '25]{The 2025 ACM Conference on Fairness, Accountability, and Transparency}{June 23--26, 2025}{Athens, Greece}
\acmBooktitle{The 2025 ACM Conference on Fairness, Accountability, and Transparency (FAccT '25), June 23--26, 2025, Athens, Greece}\acmDOI{10.1145/3715275.3732138}
\acmISBN{979-8-4007-1482-5/2025/06}

\usepackage{tabularx}
\usepackage{multirow}
\usepackage{multicol}
\usepackage{amsmath}
\newtheorem{definition}{Definition}

\usepackage{siunitx}
\usepackage{dsfont}

\usepackage{graphicx}
\usepackage{caption}
\usepackage{subcaption}
\usepackage{hyperref}
\usepackage{array}
\begin{document}

\title[Allocation Multiplicity]{Allocation Multiplicity:\\ Evaluating the Promises of the Rashomon Set}

\author{Shomik Jain}
\affiliation{
 \department{Institute for Data, Systems, and Society}
 \institution{Massachusetts Institute of Technology}
 \city{Cambridge}
 \state{MA}
 \country{USA}
 }

\author{Margaret Wang}
\affiliation{
 \department{Department of Electrical Engineering \& Computer Science}
 \institution{Massachusetts Institute of Technology}
 \city{Cambridge}
 \state{MA}
 \country{USA}
 }

\author{Kathleen Creel}
\authornote{These authors contributed equally.}
\affiliation{
 \department{Department of Philosophy \& Religion and\\ Khoury College of Computer Sciences}
 \institution{Northeastern University}
 \city{Boston}
 \state{MA}
 \country{USA}
 }

\author{Ashia Wilson}
\authornotemark[1]
\affiliation{
 \department{Department of Electrical Engineering \& Computer Science}
 \institution{Massachusetts Institute of Technology}
 \city{Cambridge}
 \state{MA}
 \country{USA}
 }

\begin{abstract}
The Rashomon set of equally-good 
models promises less discriminatory algorithms, reduced outcome homogenization, and fairer decisions through model ensembles or reconciliation. However, we argue from the perspective of \textit{allocation multiplicity} that these promises may remain unfulfilled. When there are more qualified candidates than resources available, many different allocations of scarce resources can achieve the same utility. This space of equal-utility allocations may not be faithfully reflected by the Rashomon set, as we show in a case study of healthcare allocations. We attribute these unfulfilled promises to several factors: limitations in empirical methods for sampling from the Rashomon set, the standard practice of deterministically selecting individuals with the lowest risk, and structural biases that cause all equally-good 
models to view some qualified individuals as inherently risky.
\end{abstract}





\maketitle

\section{Introduction}

Many decision-making settings involve allocating a fixed number of resources or opportunities among individuals. A company might have hundreds of job applicants but only ten job openings to fill; a healthcare program may be able to enroll only a 100 patients out of thousands. In these cases, there are often more qualified candidates than resources or goods available. This creates \textit{allocation multiplicity}: the existence of many equal-utility selections of individuals. For example, there are over 3.2 million possible ways to choose 10 qualified people from 25 qualified applicants. The number of possible equal-utility allocations increases even more if we assume that decision-makers will make mistakes and choose some unqualified individuals. 

Decision-makers often rely on AI models to determine allocations based on \textit{individual risk}. The standard practice is to select individuals with the lowest risk or highest likelihood of being qualified based on measurable features and available training data. But as several works have pointed out~\citep{black2022multiplicity, marx_predictive_2020, watson2023predictive}, individuals may receive different risk scores from models with similar training performance, a phenomenon known as \textit{model multiplicity}. The full space of models with near-equal performance is known as the \textit{Rashomon set}~\citep{breiman2001statistical, fisher2019all}, and a growing literature seeks to find as many models from this set as possible~\citep{watson2023multi, rudin2024position}. Using the Rashomon set to inform decision-making may have several benefits. One is the chance to find \textit{less discriminatory algorithms}~\citep{black2024less, gillis2024operationalizing} that improve group fairness without sacrificing accuracy. Another is its potential to reduce \textit{outcome homogenization}~\citep{bommasani2022outcome, toups2023ecosystem}, or the systemic rejection of individuals in multi-shot contexts, if decision-makers adopt different models within the set~\citep{creel_hellman_2022, jain2024facct, jain2024icml}. A third is the opportunity to use the Rashomon set to \textit{ensemble}~\citep{black2021selective,cooper2024arbitrariness} or \textit{reconcile}~\citep{du2024reconciling, roth2023reconciling} conflicting predictions.

\begin{figure*}[h!]
    \centering
    \includegraphics[width=0.75\linewidth]{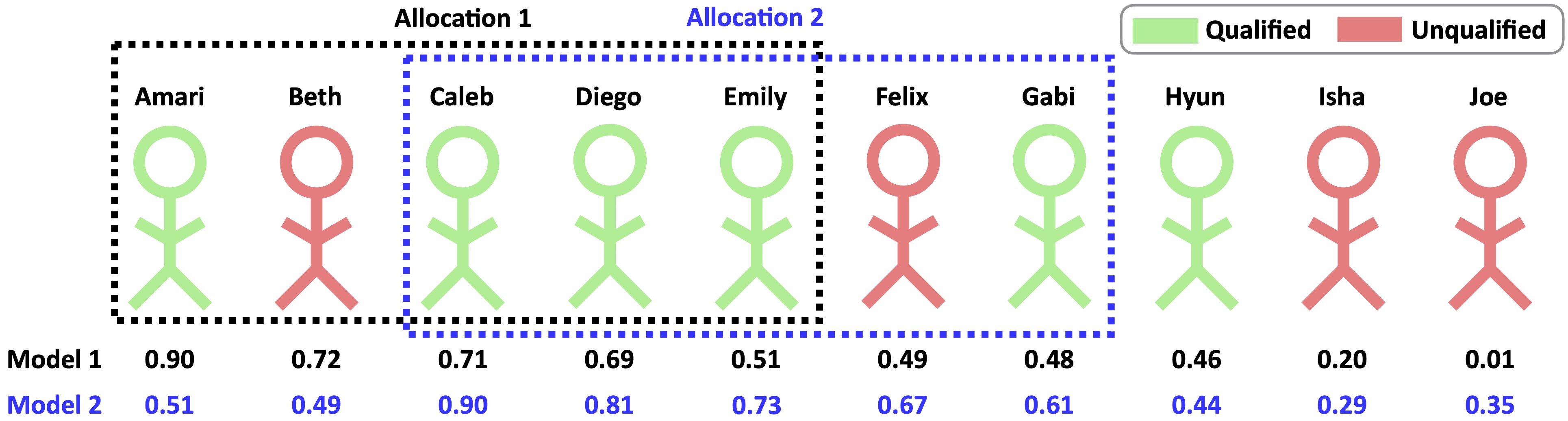}
    \caption{Suppose that 10 individuals with different qualifications compete for 5 scarce resources. 2 models from the Rashomon set yield different predictions or risk scores, leading to 2 different \textit{equal-utility} allocations in which 4 out of 5 selected individuals are qualified. In this setting, there are 60 such equal-utility allocations based on ${6 \choose 4} \cdot {4 \choose 1}$.}
    \Description{Suppose that 10 individuals with different qualifications compete for 5 scarce resources. 2 models from the Rashomon set yield different predictions or risk scores, leading to 2 different equal-utility allocations in which 4 out of 5 selected individuals are qualified. In this setting, there are 60 such equal-utility allocations based on {6 choose 4} cdot {4 choose 1}.}
    \label{fig:candidates}
\end{figure*}

\begin{figure*}[h!]
    \centering
    \includegraphics[width=0.73\linewidth]{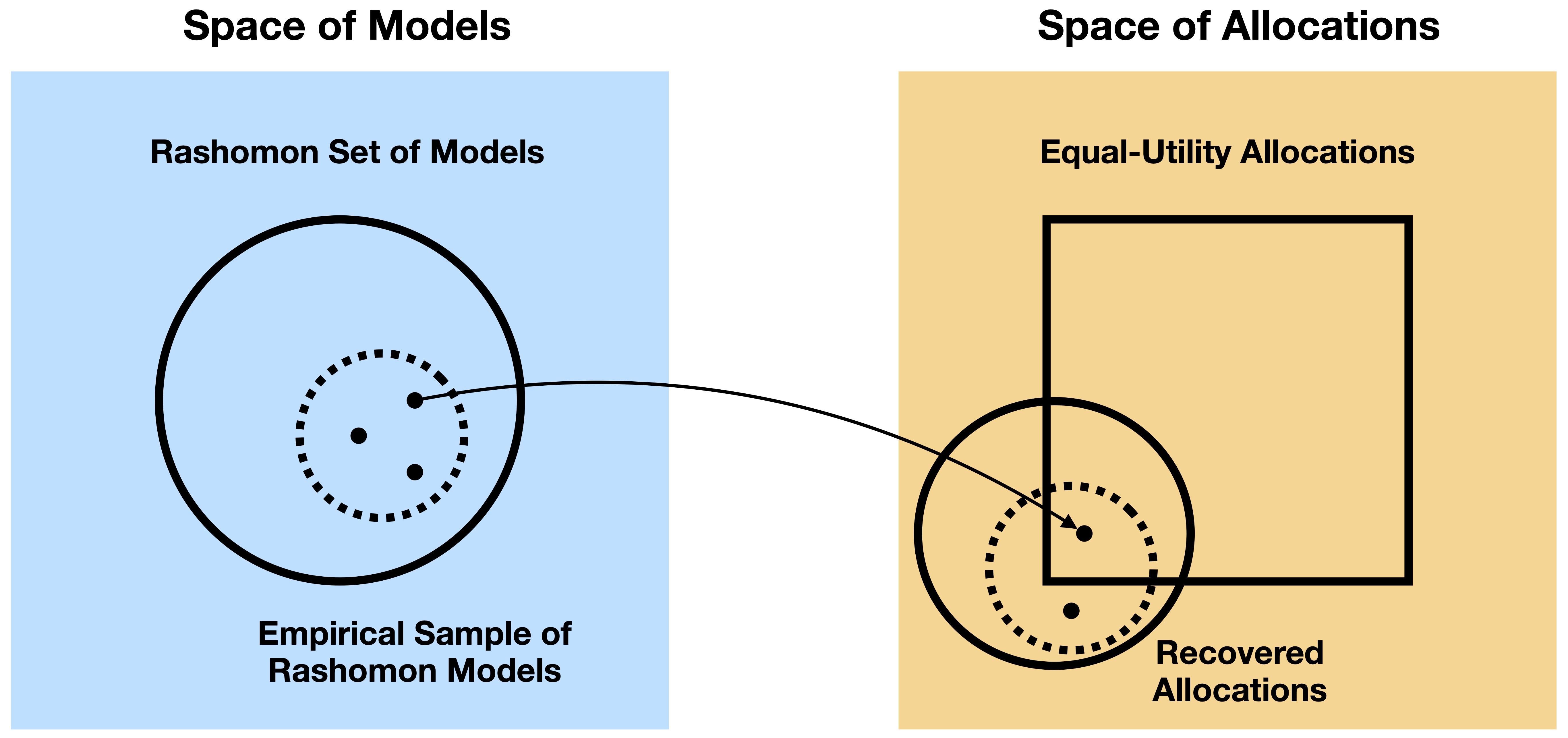}
    \caption{The Rashomon set of all equally-good models is not enumerable, but empirical methods can sample from it. This sample of Rashomon models will lead to  recovered allocations that may not faithfully represent the full space of equal-utility allocations due to (1) train-test data differences, (2) deterministic selection of the lowest risk individuals, and (3) structural biases that cause all near-optimal models to view some qualified individuals as inherently risky.}
    \Description{The Rashomon set of all equally-good models is not enumerable, but empirical methods can sample from it. This sample of Rashomon models will lead to  recovered allocations that may not faithfully represent the full space of equal-utility allocations due to (1) train-test data differences, (2) deterministic selection of the lowest risk individuals, and (3) structural biases that cause all near-optimal models to view some qualified individuals as inherently risky.}
    \label{fig:framework}
\end{figure*}

In this work, we argue that the Rashomon set primarily achieves these aims to the extent that it supports allocation multiplicity. \autoref{fig:candidates} shows that model multiplicity results in different individual-level predictions, which lead to different \textit{equal-utility allocations}. However, the Rashomon set of models will not yield the full space of equal-utility allocations for several reasons. First, it is infeasible to enumerate all possible models in the Rashomon set for unbounded hypothesis classes, so we must rely on empirical methods to sample from it~\citep{li2024practical, semenova2022existence}. This sample of Rashomon models is typically characterized by similar training or validation performance~\citep{watson2023multi, marx_predictive_2020}, which differs from deployment utility in a resource-constrained setting. Second, decision-makers usually allocate resources by deterministically selecting individuals with the lowest risk, overlooking stochastic mappings between predictions and decisions~\citep{jain2024icml}. Third, structural biases in the measurable features and available training data may cause all models in the Rashomon set to consider some qualified individuals as inherently risky. As \autoref{fig:framework} illustrates, these reasons may lead to recovered allocations from a sample of Rashomon models that are highly unrepresentative of the full space of equal-utility allocations.

The concept of allocation multiplicity enables us to evaluate how well the Rashomon set fulfills its promises of less discriminatory algorithms, reduced outcome homogenization, and fairer decisions through model ensembles or reconciliation. Through a case study of healthcare allocations, we explore various methods for sampling from the Rashomon set and compare their recovered allocations to the full space of equal-utility allocations. Our results demonstrate that: (1) many less discriminatory allocations are not found, (2) outcome homogenization persists in recovered allocations, and (3) ensemble models are unrepresentative of the full space of equal-utility allocations. These findings underscore that recovering diverse equal-utility allocations is essential to fulfilling the promises of the Rashomon set.

\section{Model Multiplicity and the Rashomon Set}
A growing literature discusses the phenomenon of \textit{model multiplicity}: the existence of many equally-good models that differ in their individual-level predictions or other properties~\citep{black2022multiplicity, marx_predictive_2020, watson2023predictive}.  \textit{Equally-good models} are models with similar performance for a chosen metric (e.g. accuracy or loss) on a fixed dataset~\citep{ganesh2024empirical, watson2023multi}. Model multiplicity can arise due to \textit{procedural multiplicity}, which refers to different choices along the model-building pipeline~\citep{black2022multiplicity, black2023pipeline}. This includes decisions about the problem formulation, feature selection, training data, and model class~\citep{watson2023multi, meyer2023dataset}. However, even when these choices are fixed, model multiplicity can still arise due to uncertainty or noise in the data as well as randomness in the model training process. For example, the data may have been generated through noisy or nondeterministic processes, such as human decision-making~\citep{rudin2024position, semenova2023path}. The data may contain insufficient information for a model to learn generalizable domain knowledge, a problem known as \textit{underspecification}~\citep{d2022underspecification}. Additionally, randomness in the model training process, such as weight initialization and random seeds, further contribute to model multiplicity~\citep{ganesh2023impact}.

The \textit{Rashomon set}\footnote{Leo \citet{breiman2001statistical} named the Rashomon set after the movie ``Rashomon,'' directed by Akira Kurosawa, in which four different perspectives on a murder fail to converge on a single truth.} consists of all models with near-optimal performance~\citep{breiman2001statistical, fisher2019all, rudin2024position}. While this set generally includes all ``good'' models, the set can also be defined for a specific hypothesis class, evaluation dataset, and loss function~\citep{hsu2022rashomon, semenova2022existence, watson2023multi}. This allows for analyses of its size through metrics like the Rashomon ratio~\citep{semenova2022existence} and Rashomon capacity~\citep{hsu2022rashomon}. Several empirical methods exist to iterate over the Rashomon set for simple hypothesis classes or sample from it for more complex model types. For example, the Rashomon set is enumerable for decision-trees with a low regularized risk value~\citep{xin2022exploring}, sparse generalized additive models~\citep{zhong2023exploring}, and sparse scoring systems~\citep{liu2022fasterrisk}. Complex hypothesis classes like neural networks have methods to explicitly search for Rashomon models, such as adversarial weight perturbation~\citep{hsu2022rashomon} and dropout-based exploration~\citep{hsu2024dropout}. Other approaches may implicitly produce Rashomon models, such as training on bootstrap datasets~\citep{cooper2024arbitrariness} or changing random seeds~\citep{ganesh2023impact}.

\subsection{Promises of the Rashomon Set}

Previous works discuss several potential benefits of model multiplicity and the Rashomon set~\citep{black2022multiplicity, rudin2024position}. Below, we highlight the benefits that are most relevant\footnote{Other potential benefits of the Rashomon set include the existence of ``simple-yet-accurate'' models for interpretability and generalizability~\citep{rudin2024position, black2022multiplicity, semenova2022existence}.} to our discussion of allocation multiplicity. We refer to these potential benefits as the ``promises'' of the Rashomon set, as they motivate research on the Rashomon set but may not be fulfilled in practice.

\subsubsection*{\textbf{Less Discriminatory Algorithms}} Model multiplicity implies that for a given model and a fairness metric, there may exist ``less discriminatory algorithms'' (LDAs) that are equally-accurate~\citep{black2022multiplicity, black2024less} but less discriminatory according to the fairness metric. Several works have shown empirically that LDAs exist for fixed populations~\citep{auerbach2024testing, gillis2024operationalizing, coston2021characterizing}, but the chosen LDA may be more discriminatory in unseen deployment settings~\citep{black2024d, laufer2024fundamental}. The goal of searching for LDAs is usually to find a model that improves on group fairness metrics like statistical parity or disparities in error rates without compromising accuracy. This motivates using the Rashomon set to find LDAs~\citep{black2022multiplicity, rudin2024position}. In domains such as credit, employment, and housing, decision-makers may have a duty to search for LDAs based on U.S. anti-discrimination laws~\cite{black2024less}. Under the \textit{disparate impact} doctrine, defendants can justify using a model with discriminatory effects by claiming the model serves a ``business necessity,'' which often relates to its performance. But even with this justification, defendants may still face liability if the plaintiffs can show there exists a ``less discriminatory alternative'' that serves the same purpose. In some cases, the courts have placed the burden on defendants to conduct a ``reasonable search'' for LDAs~\cite{black2024less, laufer2024fundamental}.

% For example, \citet{gillis2024operationalizing} operationalizes the search for LDAs across linear classifiers in the context of fair lending. \citet{coston2021characterizing} also shows that there exist LDAs compared to the COMPAS risk assessment algorithm for criminal defendants. 

\subsubsection*{\textbf{Reduced Outcome Homogenization}} Prior works have argued that model multiplicity could help mitigate homogenization across multiple decision-makers~\cite{creel_hellman_2022, jain2024facct, jain2024icml}. \textit{Outcome homogenization} occurs when the same individuals are subject to consistent errors or negative outcomes~\citep{bommasani2022outcome, toups2023ecosystem}. This represents a grave moral concern because of the possibility that individuals are \textit{systemically excluded} from important opportunities~\citep{creel_hellman_2022, jain2024facct}. Outcome homogenization often arises from \textit{algorithmic monoculture}, which occurs when decision-makers rely on identical or highly similar models~\citep{kleinberg2021algorithmic, bommasani2022outcome}. For example, job applicants may receive correlated hiring decisions across companies using similar resume-screening tools. Since the Rashomon set offers several equally-good models, it could reduce homogenization if decision-makers choose models that lead to sufficiently different outcomes for individuals~\citep{creel_hellman_2022, jain2024facct, jain2024icml}. To this end, \citet{creel_hellman_2022} suggest that decision-makers should randomly choose a model from the Rashomon set. \citet{jain2024icml} propose another approach to randomization that uses equally-good models to first quantify variance in predictions and then randomizes decisions for individuals with high variance.

\subsubsection*{\textbf{Fairer Decisions With Ensembles or Reconciliation}}

Several works use the Rashomon set to ensemble or reconcile predictions~\citep{black2021selective, cooper2024arbitrariness,roth2023reconciling,long2024individual,jiang2023recourse, du2024reconciling}. A common perspective in these works is that choosing and using a single model from the Rashomon set is ``arbitrary''\footnote{We disagree with this definition of arbitrariness and follow the definition by \citet{creel_hellman_2022}, which considers a model to be arbitrary if it doesn't serve the purposes of the decision-maker. By this definition, a model from the Rashomon set cannot be arbitrary. However, we agree that using a single model may be unfair for other reasons such as its repeated use at scale~\citep{creel_hellman_2022} or its violation of individual claims~\citep{jain2024icml}.} or unfair to individuals that could have received a more positive prediction under another Rashomon model. To address these concerns, one approach is to ensemble predictions across the Rashomon set of models~\citep{black2021selective,cooper2024arbitrariness, long2024individual, jiang2023recourse}. \citet{cooper2024arbitrariness} propose ensembles that abstain from making predictions for individuals with high inconsistencies. Another approach is to reconcile predictions ~\citep{du2024reconciling, roth2023reconciling} through a process similar to multi-calibration~\citep{hebert2018multicalibration}. Reconciliation considers two models that disagree on individual predictions and uses the disagreement region to iteratively produce a model with better empirical risk minimization.

\subsection{Related Work}

We build on several related works that evaluate the Rashomon set in deployment settings~\citep{li2024practical, laufer2024fundamental, watson2023multi, jain2024icml, simson2024one}. Our work differs in that we focus on \textit{resource-constrained allocations}. \citet{li2024practical} distinguish between the theoretical Rashomon set and the \textit{sampled Rashomon set} from empirical methods in practice. In the context of LDAs, \citet{laufer2024fundamental} highlight how near-optimal models that are less discriminatory for fixed populations may not be less discriminatory for unseen populations. They also characterize the difficulty of finding the \textit{least discriminatory algorithm}. \citet{jain2024icml} study a similar setting of scarce resource allocations, but focus on the need for randomization in this setting and do not explicitly evaluate models from the Rashomon set. Most similar to our work is that of \citet{watson2023multi}, who explore model multiplicity under resource constraints. They define the concept of \textit{top-$k$ ambiguity}, which refers to whether any model from the Rashomon set would place an individual among the top $k$ predicted values. We extend their results beyond linear models and consider top-$k$ ambiguity in the context of all possible equal-utility allocations. This space of possible decisions is similar to what \citet{simson2024one} call \textit{multiverse analysis}, although their focus is on procedural multiplicity and not a resource-constrained setting.

\section{Defining Allocation Multiplicity}

In this section, we define the concept of \textit{allocation multiplicity} and distinguish it from both model multiplicity and the Rashomon set. We also define and quantify the set of all possible \textit{equal-utility allocations}. 

\subsection{Preliminaries and Problem Formulation}

Consider a setting in which there are $n$ individuals and $k$ resources or opportunities to allocate among them. For example, a company might only have $k$ job openings or a healthcare program might only be able to enroll $k$ patients. An allocation involves the assignment of outcomes $o_i \in \{0, 1\}$ to each individual $i$. Importantly, there is scarcity in that not all individuals can receive positive outcomes with $k\ll n$.

\begin{definition}
\label{def:allocation}
An \textbf{allocation} $\mathbf{o} = (o_1, o_2, \dots, o_n)$ involves the assignment of outcomes $o_i \in \{0, 1\}$ to each individual $i$ such that only $k$ out of $n$ individuals receive positive outcomes, under a \textbf{selection rate} $\frac{k}{n}$. The space of all allocations is: $$\mathcal{O}_{k,n} = \left\{\mathbf{o} \in \{0,1\}^n\,|\,\sum^n_{i=1}o_i = k\right\}\,\text{where}\,\, |\mathcal{O}_{k,n}| = {n \choose k}$$
\end{definition}

\noindent Individuals may have varying qualifications $o^*_i$ for the resources or opportunities being allocated. We simplify to the case where each individual is qualified ($o^*_i = 1$) or unqualified ($o^*_i = 0$). For example, a patient may be qualified for enrollment in a healthcare program if they have a chronic illness. In many settings, the number of qualified individuals $n'$ is greater than the number of available resources or opportunities $k$.

\begin{definition}
\label{def:qualification_rate}
The \textbf{qualification rate} $\frac{n'}{n}$ is the number of qualified individuals $n'$ out of the total number of candidate individuals $n$.
\end{definition}

\noindent Decision-makers do not know whether individuals are qualified at the time of allocation. For many decision-makers, the objective\footnote{When $k$ is fixed, our formulation of utility is equivalent to other formulations~\citep{gillis2024operationalizing, laufer2024fundamental} that assign a benefit for selecting qualified individuals and a penalty for selecting unqualified individuals. However, we do not consider the possibility of varying benefits across qualified individuals.} is to maximize the number of qualified individuals selected in an allocation, which we denote using $k'$.

\begin{definition}
\label{def:utility}
The \textbf{utility} or \textbf{precision} of an allocation is $\frac{k'}{k} = \frac{1}{k}\sum_{i=1}^{n} o^*_i \cdot o_i$, which is simply the number of selected qualified individuals $k'$ out of the total number of selected individuals $k$.
\end{definition}

\noindent Risk-averse decision-makers aim to maximize utility by selecting individuals with the highest probability of being qualified. While the true probability $p_i = \mathbb{P}(o^*_i = 1)$ is unknowable\footnote{Individual probabilities misrepresent qualifications because the events in question are typically realized only once~\citep{dawid2017individual, dwork2021outcome, roth2023reconciling}.}, the conditional probability $p(x_i) = \mathbb{P}(o^*_i = 1\,|\,x_i)$ can be estimated using data. The validity of taking $p(x_i)$ to be an estimate of $p_i$ depends on the choice of features and how well qualifications can be predicted from them. Setting these specifications aside, machine learning models\footnote{Our focus is on non-causal predictions of individual risk or utility because many decision-makers allocate resources in this way~\citep{barocas2023fairness}. However, allocation multiplicity may still arise in a causal setting, if many individuals have similar treatment effects.} can estimate $p(x_i)$ based on features $x_i$ from a given domain $\mathcal{X}$.

\begin{definition}
\label{def:model}
A \textbf{model} $\mathbf{\hat{p}}: \mathcal{X} \rightarrow [0,1]$ maps an individual's features $x_i \in \mathcal{X}$ to a prediction $\hat{p}(x_i)$, which estimates the conditional probability $p(x_i) = \mathbb{P}(o^*_i = 1\,|\,x_i)$. 
\end{definition}

\noindent In data-driven allocations, outcomes $o_i$ are assigned based on predictions $\hat{p}(x_i)$. Standard practice in machine learning is to deterministically assign $o_i = 1$ to individuals with the top $k$ predictions. However, there are various other ways to map predictions to outcomes. For example, a human-in-the-loop could determine outcomes for high uncertainty predictions~\citep{cooper2024arbitrariness}, or a weighted lottery based on predictions could determine the outcomes~\citep{jain2024icml}.

\begin{definition}
\label{def:mapping}
A \textbf{mapping} $\mathcal{M} = \{(\hat{p}(x_1), o_1), (\hat{p}(x_2), o_2),\,\dots$\\$(\hat{p}(x_n), o_n)\}$ characterizes the relationship between a model $\mathbf{\hat{p}}$ and allocation $\mathbf{o}$. In a \textbf{top $k$ mapping}, $o_i = 1$ 
if and only if $\,\hat{p}(x_i)$ is in the $k$ highest predictions.
\end{definition}

\subsection{From Model Multiplicity to Allocation Multiplicity}\label{subsec:mm_to_am}

Model multiplicity refers to the existence of multiple models with similar performance but different properties, including producing different predictions for certain individuals (a subtype of model multiplicity called ``predictive multiplicity'') ~\citep{marx_predictive_2020, black2022multiplicity, watson2023predictive}. Performance is measured on a training or validation dataset $\mathcal{D_{\texttt{train}}} = \{(x_i, o^*_i)\}^{n_{\texttt{train}}}_{i=1}$ because the target variable $o^*_i$ (in our case, qualification) is not known at the time of allocation. Following definitions of model multiplicity in prior work~\citep{marx_predictive_2020, watson2023multi}, we consider a baseline model $\mathbf{\hat{p}_0}$ that is the solution to an empirical risk minimization problem of the form $\min_{\mathbf{\hat{p}}\in\mathcal{H}} L(\mathbf{\hat{p}}; \mathcal{D_{\texttt{train}}})$, over a hypothesis class $\mathcal{H}$ with loss function $L(\cdot\,;\mathcal{D})$. For example, $\mathcal{H}$ could be the set of neural networks represented by a specific architecture and $L(\cdot\,;\mathcal{D})$ could be the cross-entropy loss. In this context, the $\epsilon-$Rashomon set encompasses all models that achieve near-optimal loss.

\begin{definition}
\label{def:rashomonmodels}
For a baseline model $\mathbf{\hat{p}_0}$ and loss tolerance $\epsilon > 0$, the \textbf{$\mathbf{\epsilon}$-Rashomon set of models} is:
$$\mathcal{R}_\epsilon(\mathbf{\hat{p}_0}) := \{\mathbf{\hat{p}} \in \mathcal{H}: L(\mathbf{\hat{p}}; \mathcal{D_{\texttt{train}}}) \leq L(\mathbf{\hat{p}_0}; \mathcal{D_{\texttt{train}}}) + \epsilon\}$$
and \textbf{model multiplicity} exists if $\,|\mathcal{R}_\epsilon(\mathbf{\hat{p}_0})| > 1$. An \textbf{$\mathbf{\epsilon}$-Rashomon model} is a model $\mathbf{\hat{p}} \in \mathcal{R}_\epsilon(\mathbf{\hat{p}_0})$.
\end{definition}

\noindent Building on the concept of model multiplicity, we define \textit{allocation multiplicity} as the existence of multiple allocations with near-equal deployment utility. Recall that an allocation $\mathbf{o}$ is the assignment of outcomes to individuals (Definition~\ref{def:allocation}) and that its utility depends on $k'$: the number of selected individuals who are qualified  (Definition~\ref{def:utility}). Let $\mathbf{o_0}$ denote an allocation based on a mapping  $\mathcal{M}_0$ from the baseline model's predictions $\mathbf{\hat{p}_0}$. In this context, the set of $\Delta$-equal-utility allocations encompasses all allocations that achieve near-equal utility.

\begin{definition}
\label{def:rashomonallocations}
For a baseline allocation $\mathbf{o_0}$ and utility tolerance $\Delta \geq 0$, the \textbf{set of $\Delta$-equal-utility allocations} is:
$$\mathcal{R}_\Delta(\mathbf{o_0}) := \left\{\mathbf{o_\delta} \in \mathcal{O}_{k,n} \,|\, k'_\delta \geq k'_0 - \Delta\right\}$$
and \textbf{allocation multiplicity} exists if $\,|\mathcal{R}_\Delta(\mathbf{o_0})| > 1$.  A \textbf{$\mathbf{\Delta}$-equal-utility allocation} is an allocation $\mathbf{o} \in \mathcal{R}_\Delta(\mathbf{o_0})$. When $\Delta=0$, $\mathcal{R}_0(\mathbf{o_0})$ is a set of \textbf{equal-utility allocations}.
\end{definition}

\noindent An $\epsilon$-Rashomon model does not guarantee a $\Delta$-equal-utility allocation for several reasons. First, $\epsilon$ measures deviation from the optimal training or validation loss, while $\Delta$ measures deviation from a specific deployment utility. These metrics are not directly comparable given that the loss is not evaluated in a resource-constrained setting~\citep{watson2023multi, jain2024icml}. Generalization error adds further uncertainty, as training or validation performance might differ from deployment utility. Moreover, predictions can be mapped to outcomes in various ways, such as through stochastic procedures or with a human-in-the-loop. Thus, even a model outside the $\epsilon$-Rashomon set could yield a $\Delta$-equal-utility allocation. For example, a human decision-maker might have domain knowledge for high-uncertainty predictions, helping them decide on a high-utility allocation when using a sub-optimal model. Even for the top $k$ mapping, a model could make predictions with high loss (e.g. predictions clustered around 0.5), yet maintain an ordering that yields high utility.

\subsection{Quantifying Allocation Multiplicity}

The number of equal-utility allocations is often exponentially high. We can calculate how many equal-utility allocations are possible based on the number of total individuals $n$, qualified individuals $n'$, resources $k$, and qualified individuals selected $k'$ under the baseline allocation $\mathbf{o}_0$. An allocation will have the same utility as $\mathbf{o}_0$ if it selects exactly $k'$ qualified individuals and $k-k'$ unqualified individuals. This represents a combinatorial problem, as there are $n'$ qualified individuals to choose $k'$ from, and $n-n'$ unqualified individuals to choose $k-k'$ from. 

\begin{proposition}
\label{prop:num_allocations}
For a setting in which there are $n$ individuals, $k$ resources, $n'$ qualified individuals, and $k'$ qualified individuals selected by a baseline allocation $\mathbf{o}_0$, the \textbf{number of $\mathbf{\Delta}$-equal-utility allocations} is:
$$|\mathcal{R}_\Delta(\mathbf{o}_0)| = \sum_{k'_\delta\,=\,k' - \Delta}^{k'} {n' \choose k'_\delta} \cdot {n - n' \choose k - k'_\delta}$$
\end{proposition}

\begin{table*}[t!]
\small
\centering
\begin{tabular}{cccccccc} 
\multicolumn{8}{c}{\textbf{Number of Equal-Utility Allocations} $|\mathcal{R}_0(\mathbf{o}_0)|$} \\
\addlinespace
\toprule
\multirow{2.5}{*}{$k'/k$} & \multirow{2.5}{*}{$n'/n$} & & $n=100$ & &  & $n=1000$ & \\
\cmidrule(lr){3-5} \cmidrule(lr){6-8}
& & $k/n = 0.10$ & $k/n = 0.25$ & $k/n=0.50$ & $k/n = 0.10$ & $k/n = 0.25$ & $k/n=0.50$ \\
\toprule
\addlinespace
%0.75 & 0.50   & \num{2e12} & \num{2e21} & \num{1e23} & \num{3e132} & \num{1e223} & \num{3e241} \\ 
0.85 & 0.50 & \num{7e11} & \num{2e19} & \num{3e17} & \num{9e125} & \num{6e203} & \num{9e180} \\ 
0.95 & 0.50  & \num{1e11} & \num{1e17} & \num{4e08} & \num{4e115} & \num{1e174} & \num{1e84} \\ 
%\addlinespace
%0.75 & 0.75  & \num{5e12} & \num{5e22} & \num{2e28} & \num{6e138} & \num{3e241} & \num{2e298} \\ 
0.85 & 0.75 & \num{5e12} & \num{3e22} & \num{2e27} & \num{3e137} & \num{4e237} & \num{3e286} \\ 
0.95 & 0.75 & \num{3e12} & \num{4e21} & \num{7e23} & \num{2e132} & \num{8e222} & \num{6e246} \\ 
\addlinespace
\bottomrule
\addlinespace
\end{tabular}
\caption{$|\mathcal{R}_0(\mathbf{o}_0)|$ for a baseline allocation $o_0$ with varying utility $k'/k$, qualification rate $n'/n$, and selection rate $k/n$.}
\label{tab:multiplicity}
\end{table*}

\noindent Table~\ref{tab:multiplicity} computes the number of equal-utility allocations for a baseline allocation with varying utility $k'/k$, qualification rate $n'/n$, and selection rate $k/n$. Even when there are only 100 individuals, the number of equal-utility allocations is exponentially high. For example, there are \num{2e19} equal-utility allocations with utility 0.85, qualification rate 0.50, and selection rate 0.25. This represents the number of ways to select 25 people from 100 people, where 50 people are qualified and 21 qualified people are selected. Intuitively, the number of equal-utility allocations is maximized when both: (1) the number of qualified people selected is around half the number of qualified people in the population, and (2) the number of unqualified people selected is around half the number of unqualified people in the population. This follows from the fact that binomial coefficients of the form $n \choose k$ are maximized when $k \approx \frac{n}{2}$.

\section{Why Does Allocation Multiplicity Matter?}

Now that we have seen how many equal-utility allocations are possible, we discuss why allocation multiplicity matters. First, risk-averse decision-makers might assume that all or most utility-maximizing allocations can only be produced by top $k$ mappings from near-optimal models. Allocation multiplicity highlights the existence of ``riskier'' allocations that could have been chosen without sacrificing utility. Second, decision-makers who prioritize normative ideals other than utility may value allocation multiplicity for reasons aligned with their principles. Third, those who view the Rashomon set positively for its potential benefits may find that the space of all equal-utility allocations is what actually delivers those benefits.

\subsection{From the Perspective of Normative Ideals}

\subsubsection*{\textbf{From the Perspective of Risk-Aversion}} Risk-averse decision-makers will usually choose the allocation with the highest \textit{expected} utility based on a model's predictions. From the perspective of the information they have at the time, this might be the most utility-maximizing step they can take. But model multiplicity means that even holding fixed  a problem formulation and a training dataset, there are many models that minimize risk to an equivalent extent, thereby delivering  equivalent expected utility allocations for the risk averse. However, \textit{realized} utility, not expected utility, is what ultimately matters to most decision-makers. While it may be difficult to recover all allocations with the same realized utility based on currently available information, the knowledge that additional allocations exist should spur decision-makers to collect the information needed to recover more and sample better from the space of equal-utility allocations. Using a validation dataset to compare the highest-expected-utility allocations with all allocations yielding the same realized utility provides an evaluation framework that may uncover shortcomings of the current training data and inherent limitations of allocating solely on expected risk minimization.

\begin{figure*}[t!]
\begin{subfigure}{.45\textwidth}
  \centering
  % include first image
  \includegraphics[width=\linewidth]{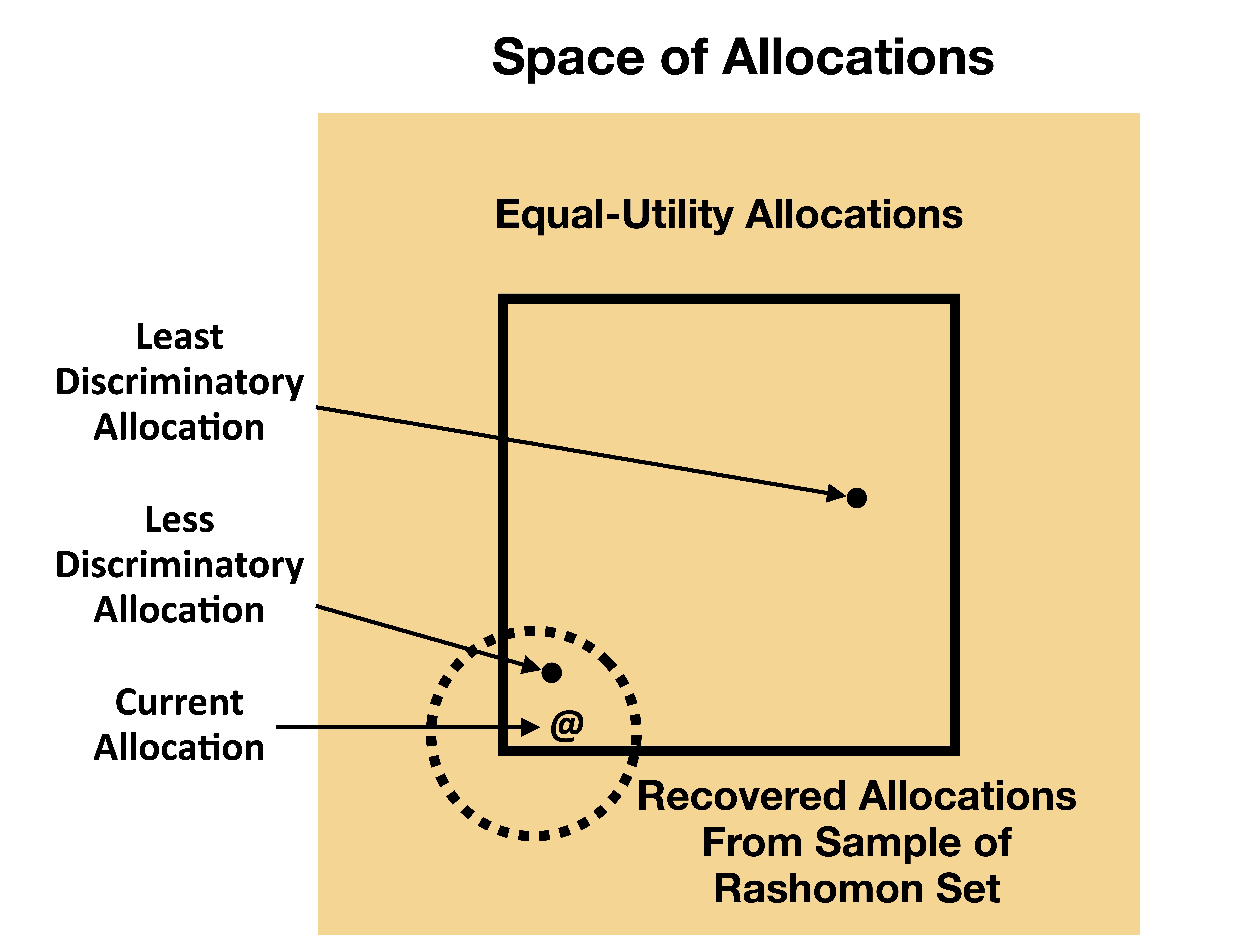}  
  \caption{Less Discriminatory Algorithms}
  \label{fig:why_lda}
\end{subfigure}
\hfill  
\begin{subfigure}{.45\textwidth}
  \centering
  % include second image
\includegraphics[width=\linewidth]{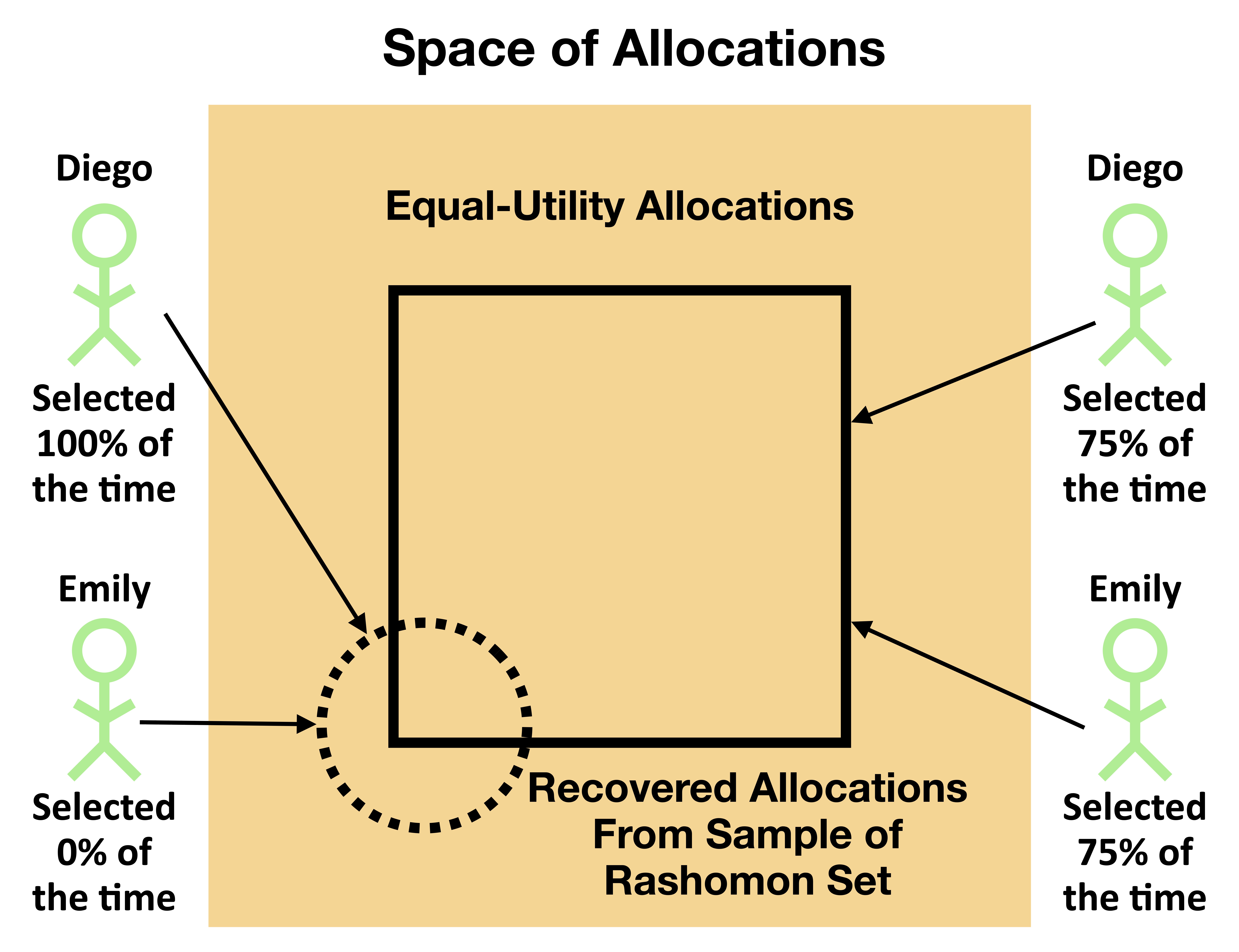}  
  \caption{Outcome Homogenization}
  \label{fig:why_homogenization}
\end{subfigure}
\caption{Evaluating Promises of the Rashomon Set from the Perspective of Allocation Multiplicity: (a) the least discriminatory equal-utility allocation may not be recovered; (b) outcome homogenization may persist in recovered allocations.}
\Description{Evaluating Promises of the Rashomon Set from the Perspective of Allocation Multiplicity: (a) the least discriminatory equal-utility allocation may not be recovered; (b) outcome homogenization may persist in recovered allocations.}
\label{fig:why_matters}
\end{figure*}

\subsubsection*{\textbf{From the Perspective of Claims}}

Allocation multiplicity has a positive role in ensuring that the use of machine
learning models to make decisions lives up to our normative commitments.
For example, some decision-makers are morally committed to fulfilling the \textit{claims}~\cite{broome1990fairness} that individuals have to goods or outcomes for reasons of fairness. According to the claims
view, an individual has a claim to an organ if she \textit{needs} it more than others; a claim to a job if she \textit{merits} it; or a claim to a good if she was promised it and therefore \textit{deserves} it. Claims should be satisfied in proportion to their strength~\cite{broome1990fairness}. For example, if claims are based on qualification, then all equally qualified individuals should have an equal chance of being selected. 
Comparing recovered allocations from the Rashomon set to the space of all equal-utility allocations shows that we over-respect the claims of those who appear less risky and under-respect the claims of those who appear riskier. Although we may not always have the information needed to satisfy claims in proportion to their strength, the claims framework suggests that we should (1) seek out additional information that would allow us to better respect all claims, and (2) be prepared to give ``compensatory'' goods to those whose claims we failed to respect.

\subsubsection*{\textbf{From the Perspective of the Worst-Off}}
Prioritarianism holds that when allocating resources, we should prioritize improvements to the well-being of the worst-off~\cite{Arneson2022}. Within the space of equal-utility allocations, some allocations will give more resources to those who suffer the most. Thus, the prioritarian may want to choose an equal-utility allocation that prioritizes the worst-off, similar to the perspective of choosing the ``least discriminatory'' algorithm among a set of equally-good models~\cite{black2024less}. However, the shift in perspective from model multiplicity to allocation multiplicity suggests that many of the ``multiple'' models have allocations that are concentrated on the least-risky individuals. These individuals are likely to be the best-off, not the worst-off. Prioritarianism would suggest going beyond the Rashomon set to find the equal-utility allocations that choose the seemingly-risky people who would be most benefited by receiving positive outcomes.

\subsection{From the Perspective of the Rashomon Set}
Prior works suggest that the Rashomon set promises benefits such as less discriminatory algorithms, reduced outcome homogenization, and fairer decisions through ensembles or reconciliation. In this section, we argue that the space of equal-utility allocations is what actually promises these benefits. The extent to which the Rashomon set fulfills these promises depends on how effectively it recovers diverse equal-utility allocations. 
 
\subsubsection*{\textbf{Less Discriminatory Algorithms}}

For a given fairness metric, there exists a \textit{least discriminatory} allocation within the space of all equal-utility allocations. For example, decision-makers may accidentally under-select individuals who are racial minorities because they are perceived as riskier by predictive models. However, excluding these individuals is not necessary: provided there are at least $k'$ qualified individuals within the minority population, an equal-utility allocation exists that consists entirely of minority individuals. But since most of these allocations are difficult for predictive models to find, this may cause all ``less discriminatory algorithms'' (LDAs) in the Rashomon set to result in allocations that deviate significantly from the ``least discriminatory'' equal-utility allocation, as Figure~\ref{fig:why_lda} illustrates. Moreover, train-test data differences may further contribute to this issue: prior research has shown that LDAs derived from training or validation data can become more discriminatory when deployed in unseen deployment settings~\citep{black2024d, laufer2024fundamental}.

\subsubsection*{\textbf{Reduced Outcome Homogenization}} 
In the space of all equal-utility allocations, {\em every} individual has a non-zero probability of being selected when allocations do not achieve perfect utility. This means that the \textit{systemic rejection rate} could theoretically be reduced to zero without sacrificing utility, provided there are enough decision-makers. However, as illustrated in Figure~\ref{fig:why_homogenization}, allocations derived from the Rashomon set may exclude some qualified individuals entirely, leaving them with zero chance of selection. This underscores that predictive multiplicity, as captured by the Rashomon set, does not fully deliver on its promise of reducing homogenization. Viewing the problem through the lens of allocation multiplicity makes this limitation clear and highlights how considering the space of equal-utility allocations can provide better insights into pathways for reducing outcome homogenization.

\subsubsection*{\textbf{Fairer Decisions With Ensembles or Reconciliation}} 

Ensembling or reconciling predictions across models in the Rashomon set is often proposed as a way to achieve fairer decisions. However, these methods are limited when viewed through the lens of allocation multiplicity. While they resolve inconsistent predictions across models, the aggregated predictions may not capture the diversity of equal-utility allocations that the Rashomon set implicitly supports. Ensembling or reconciling predictions is necessarily based on the starting set of Rashomon predictions, but allocations yielded by the Rashomon set (via a top-$k$ mapping) may not form an even representation of the set of equal-utility allocations. For example, the average characteristics of individuals chosen across the space of equal-utility allocations may differ greatly from the average characteristics of individuals chosen in allocations recovered by the Rashomon set. This limitation underscores the need to consider the space of equal-utility allocations directly, at it provides a more comprehensive view of fairness by encompassing all diverse outcomes.

\section{Case Study: Healthcare Allocations}

We turn to a case study of healthcare allocations in order to analyze allocation multiplicity in a real-world setting. Specifically, we evaluate how well the Rashomon set fulfills its promises from the perspective of allocation multiplicity. Our results show that: (1) many less discriminatory allocations are not found, (2) outcome homogenization persists in recovered allocations, and (3) the allocation from an ensemble model is unrepresentative of the full space of equal-utility allocations.

\subsection{Data and Methods}

We use a healthcare dataset\footnote{\citet{obermeyer2019dissecting} compiles data for all primary care patients at a large academic hospital. Due to the sensitivity of this data, the study authors released a semi-synthetic version that is designed to closely mirror the original dataset. This data is available at: \url{https://gitlab.com/labsysmed/dissecting-bias}.} released by \citet{obermeyer2019dissecting} that was used to select patients for a ``high-risk care management'' program at a hospital. The hospital aimed to identify high-risk patients in a given year based on information about the patient's health and healthcare costs in the previous year. Specifically, the dataset contains features\footnote{In our experiments, we use the same subset of features as \citet{watson2023predictive} (see Appendix Table~\ref{tab:features}). In particular, we do not use race as a feature.} for demographics (age, race, sex), chronic illnesses that a patient had in the previous year, and costs claimed by the patients' insurer in the previous year. These costs are broken down for various healthcare services, such as primary care, emergency room visits, and pharmacy expenses.

\begin{table*}[t!]
\small
\centering
\begin{tabular}{cccccccccccc} 
\toprule
& & \multicolumn{2}{c}{Different} & \multicolumn{2}{c}{Bootstrapped}  & \multicolumn{2}{c}{Shuffled}  & \multicolumn{2}{c}{Weight} & \multicolumn{2}{c}{Decision-Boundary}\\
& & \multicolumn{2}{c}{Feature Subsets} & \multicolumn{2}{c}{Training Data}  & \multicolumn{2}{c}{Data Order}  & \multicolumn{2}{c}{Perturbation} & \multicolumn{2}{c}{Randomization}\\
\cmidrule(lr){2-2} \cmidrule(lr){3-4} \cmidrule(lr){5-6} \cmidrule(lr){7-8} \cmidrule(lr){9-10} \cmidrule(lr){11-12}
& $\frac{n'}{n} $& $\frac{k}{n}=0.25$ & $\frac{k}{n}=0.50$ & $\frac{k}{n}=0.25$ & $\frac{k}{n}=0.50$ & $\frac{k}{n}=0.25$ & $\frac{k}{n}=0.50$ & $\frac{k}{n}=0.25$ & $\frac{k}{n}=0.50$ & $\frac{k}{n}=0.25$ & $\frac{k}{n}=0.50$\\
\addlinespace
\toprule
\multirow{2}{*}{\shortstack{Number of\\Rashomon Models}}    & 0.32 & 1674 & 1674 & 997 & 997 & 959 & 959 & 999 & 999 & 1 & 1 \\
 & 0.55 & 1386 & 1386 & 1000 & 1000 & 835 & 835 & 1000 & 1000 & 1 & 1 \\
\midrule
\multirow{2}{*}{\shortstack{Validation\\Loss}}  & 0.32 & 0.279 & 0.279 & 0.284 & 0.284 & 0.284 & 0.284 & 0.290 & 0.290 & 0.281 & 0.281 \\
 & 0.55 & 0.327 & 0.327 & 0.328 & 0.328 & 0.330 & 0.330 & 0.335 & 0.335 & 0.325 & 0.325 \\
\midrule
\multirow{2}{*}{\shortstack{Number of\\Recovered Allocations}} & 0.32 & 502 & 373 & 997 & 997 & 75 & 62 & 54 & 68 & 1000 & 1000 \\
 & 0.55 & 734 & 460 & 1000 & 1000 & 184 & 74 & 26 & 31 & 1000 & 1000 \\
\midrule
\multirow{2}{*}{\shortstack{Allocation\\Utility ($k'/k$)}} 
 & 0.32 & 0.873 & 0.602 & 0.877 & 0.604 & 0.877 & 0.605 & 0.877 & 0.604 & 0.869 & 0.599 \\
 & 0.55 & 0.950 & 0.902 & 0.953 & 0.904 & 0.952 & 0.905 & 0.952 & 0.905 & 0.952 & 0.885\\
\bottomrule
\addlinespace
\end{tabular} 
\caption{\textbf{Performance Metrics For Different Methods of Sampling From the Rashomon Set}. All methods produce many models with near-equal performance, but some methods yield more unique allocations (based on a top $k$ mapping) than others.}
\label{tab:perf}
\end{table*}

Using this data, we run various simulations\footnote{The code for our experiments is available at: \url{https://github.com/mqwangwa/allocation_multiplicity}.} of a scenario in which a hospital is allocating enrollment in a healthcare program among its patients. In order to determine who is qualified for the program~\citep{obermeyer2019dissecting, watson2023multi}, the hospital could predict which patients will have the highest future healthcare costs or the most chronic illnesses. However, \citet{obermeyer2019dissecting} demonstrate that using predicted healthcare costs to select patients can result in racial bias. We therefore consider an individual to be qualified if they \textit{will} have $q$ or more chronic illnesses in the upcoming year. To test different qualification rates, we vary $q$ to represent having at least 1, 2, or 3 chronic illnesses, which corresponds to 55\%, 32\%, and 19\% of the population, respectively. Not all qualified patients can be selected because, as in the real-world,\footnote{In the real-world hospital studied by \citet{obermeyer2019dissecting}, patients with risk scores in the top 3\% were automatically enrolled in the program, while those in the top 45\% were flagged for their doctors, who could enroll them based on program availability.}  healthcare programs are resource-constrained~\citep{obermeyer2019dissecting, watson2023multi}. To simulate this, we consider different selection rates of 10\%, 25\%, and 50\%. For each combination of qualification and selection rates, we run 250 simulations\footnote{We use 10 different partitions of a 60-20-20 train-validation-test split. For each partition, we run 25 simulations where a different set of $n=1000$ patients are drawn from the test set to be candidates for our hypothetical allocation. Results with standard deviation over these 250 iterations are reported in the Appendix.} with different test sets of $n=1000$ patients as allocation candidates.

In each simulation, we consider the Rashomon set of models that predict whether individuals are qualified. Since this set is non-enumerable, we evaluate various empirical methods for sampling from the Rashomon set. We choose these methods because they represent different points along the ML pipeline that may lead to model multiplicity~\citep{black2023pipeline, black2022multiplicity}.
\begin{itemize}
    \item \textbf{Different Feature Subsets}: We follow the procedure in \citet{liu2022fasterrisk} to explore different sparse scoring systems (FasterRisk). These are linear models with integer coefficients and a restricted number of features.  
    \item \textbf{Bootstrapped Training Data}: We follow the procedure in \citet{cooper2024arbitrariness} to train different neural networks based on bootstrapped sub-samples of the training data.
    \item \textbf{Shuffled Data Order}: We follow the procedure in \citet{ganesh2023impact} to shuffle the data order between neural network training epochs, and take the predictions at each epoch to be a different ``model.'' This is an efficient way to approximate prediction variance from different weight initializations~\citep{ganesh2023impact}.
    \item \textbf{Weight Perturbation}: We follow the procedure in \citet{hsu2022rashomon} to adversarially perturb neural network weights. Perturbed models are obtained by fine-tuning to increase the predicted score for different validation data points until the validation loss exceeds $\epsilon$. 
\end{itemize}

\noindent For each method, we consider the best-performing model in terms of cross-entropy loss on a validation dataset. We define the \textbf{\textit{empirical sample of Rashomon models}} to be all models with a validation loss within $\epsilon=0.01$ of the best-performing model for a given method\footnote{The theoretical Rashomon set can vary across methods, as it is defined based on the best-performing model for that method. We choose this approach because the Rashomon set is often defined for a specific hypothesis class~\citep{watson2023multi, marx_predictive_2020}. However, in our simulations, all the methods yield models with an average validation loss around 0.01 of each other, as Table~\ref{tab:perf} shows.}. We assume the \textbf{\textit{recovered allocations}} from this sample are determined using a top $k$ mapping, where individuals with the highest $k$ predictions are selected. For each method, Table~\ref{tab:perf} shows the number of models found in the Rashomon set and the corresponding number of recovered allocations. Note that many models (unique sets of predictions) map to the same allocation based on the top $k$ mapping. Table~\ref{tab:perf} also shows the average \textbf{\textit{utility}} of recovered allocations, which is the proportion of selected patients with $q$ or more chronic illnesses. This average utility serves as the basis for defining our reference of all \textbf{\textit{equal-utility allocations}}.

For comparison, we also evaluate stochastically generated allocations using a single model's predictions and \textbf{\textit{decision-boundary randomization}}. We follow the procedure in \citet{jain2024icml} to generate allocations using a partial weighted lottery\footnote{In the Appendix, we try other parameters for decision-boundary randomization as well as another method to randomize allocations (see Table~\ref{tab:all_random_results}). Generally, randomized methods achieve better results in fulfilling the promises of the Rashomon set while still attaining comparable utility.}: the first $0.75k$ resources are deterministically given to patients with the highest predictions and then the remaining $0.25k$ resources are randomized over patients with the next $0.50k$-highest predictions. While these allocations are not recovered from a Rashomon set of models, they represent another way to find (close to) equal-utility allocations. Notably, the mean and variance of deployment utility are similar for recovered allocations from the Rashomon set and those from decision-boundary randomization (Appendix Table~\ref{tab:all_random_results}).

\begin{figure*}[t!]
\begin{subfigure}{.75\textwidth}
  \centering
  % include first image
  \includegraphics[width=\linewidth]{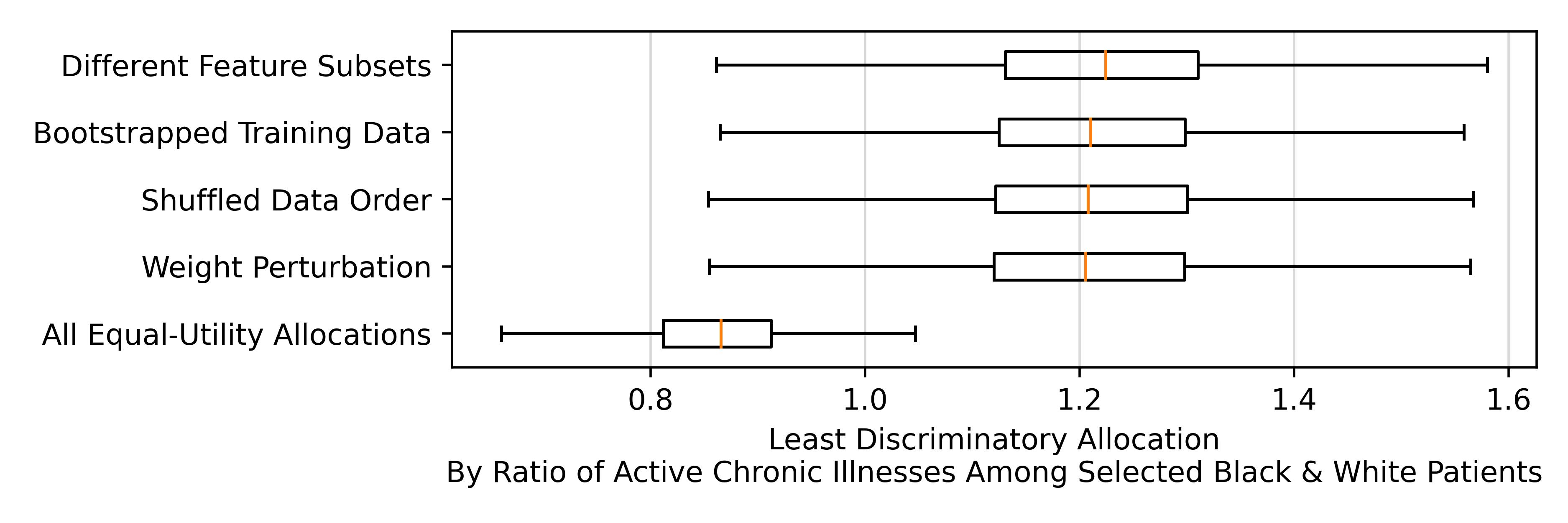}  
  \caption{The ``least discriminatory'' allocation is not found, according to the threshold test and reference equal-utility allocation that selects patients in descending order by number of chronic illnesses.}
  \label{fig:results_lda}
\end{subfigure}
\begin{subfigure}{.45\textwidth}
  \centering
  % include first image
  \includegraphics[width=\linewidth]{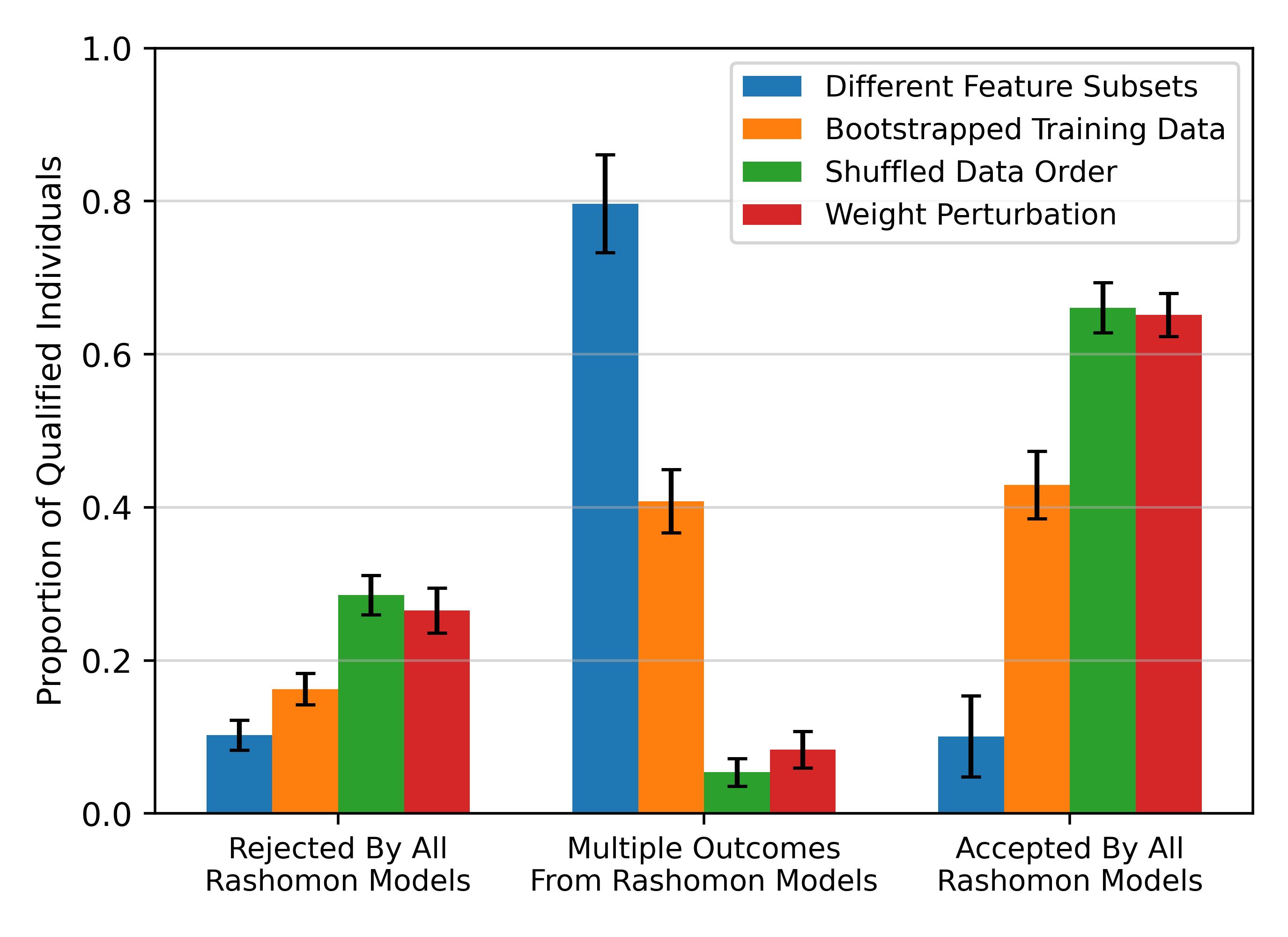}  
  \caption{Outcome homogenization persists in recovered allocations from the Rashomon set: between 10\% to 30\% of qualified individuals are rejected by all Rashomon models.}
  \label{fig:results_oh}
\end{subfigure}
\hfill
\begin{subfigure}{.45\textwidth}
  \centering
  % include second image
  \includegraphics[width=\linewidth]{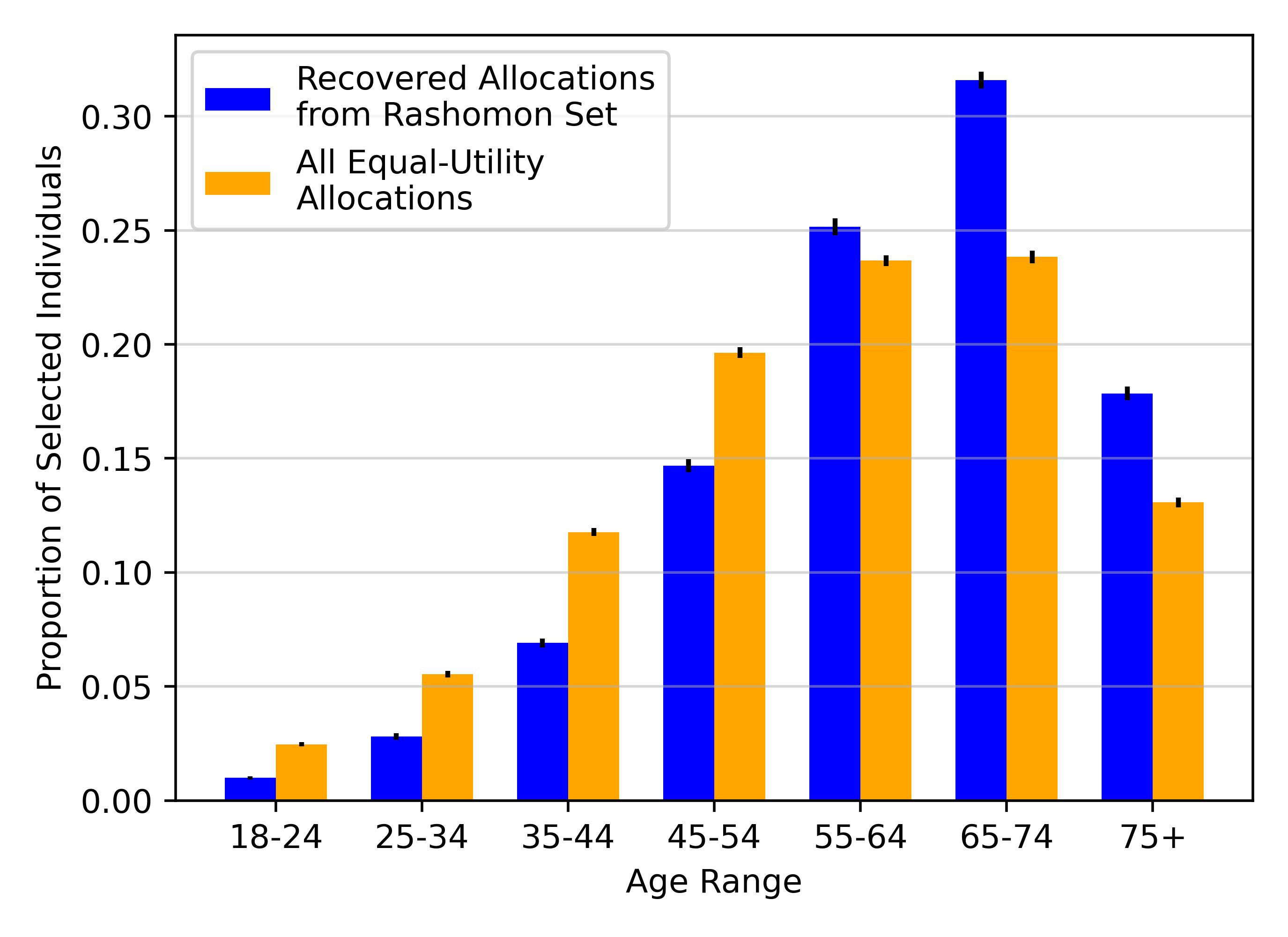}  
  \caption{The age of selected patients is skewed older on average in recovered allocation from the Rashomon set, indicating how ensembling predictions may lead to unrepresentative allocations.}
  \label{fig:results_ensemble}
\end{subfigure}
\caption{\textbf{Comparing Recovered Allocations From the Rashomon Set to All Equal-Utility Allocations ($\frac{k}{n} = 0.25$, $\frac{n'}{n}=0.32$)}}
\label{fig:results}
\end{figure*}

\begin{table*}[t!]
\small
\centering
\begin{tabular}{ccccccccc} 
\toprule
& \multirow{2}{*}{$k/n$} & \multirow{2}{*}{$n'/n$} & All Equal-Utility & Different & Bootstrapped & Shuffled & Weight & Decision-Boundary \\
& & & Allocations & Feature Subsets & Training Data & Data Order & Perturbation & Randomization \\
\toprule
\multirow{4}{*}{\shortstack{Least\\Discriminatory\\Allocation $\dagger$}}  & 0.25 & 0.32 & 0.86 & 1.12 & 1.12 & 1.19 & 1.18 & 1.11 \\
 & 0.25 & 0.55 & 0.92 & 1.10 & 1.12 & 1.18 & 1.20 & 1.14 \\
 & 0.50 & 0.32 & 1.21 & 1.26 & 1.25 & 1.31 & 1.28 & 1.23 \\
 & 0.50 & 0.55 & 0.90 & 1.25 & 1.24 & 1.30 & 1.29 & 1.23 \\

\midrule
\multirow{4}{*}{\shortstack{Outcome\\Homogenization $\ddagger$}}  & 0.25 & 0.32 & 0.80 & 0.95 & 0.97 & 0.99 & 0.99 & 0.94 \\
 & 0.25 & 0.55 & 0.71 & 0.92 & 0.95 & 0.98 & 0.99 & 0.94 \\
 & 0.50 & 0.32 & 0.69 & 0.95 & 0.96 & 0.99 & 0.98 & 0.89 \\
 & 0.50 & 0.55 & 0.76 & 0.94 & 0.96 & 0.99 & 0.99 & 0.89 \\

\midrule
\multirow{4}{*}{\shortstack{Age\\Homogenization $\star$}}   & 0.25 & 0.32 & 2.49 & 2.35 & 2.33 & 2.36 & 2.36 & 2.38 \\
 & 0.25 & 0.55 & 2.56 & 2.20 & 2.21 & 2.31 & 2.33 & 2.23 \\
 & 0.50 & 0.32 & 2.52 & 2.51 & 2.48 & 2.47 & 2.47 & 2.52 \\
 & 0.50 & 0.55 & 2.55 & 2.49 & 2.46 & 2.45 & 2.45 & 2.50 \\

\bottomrule
\addlinespace
\multicolumn{9}{l}{$\dagger$ The least discriminatory allocation is measured by the lowest found ratio of active chronic illnesses between selected Black and White patients.} \\
\multicolumn{9}{l}{$\ddagger$ Outcome homogenization is measured by pairwise consistency: the probability of an individual receiving the same outcome in 2 random allocations.} \\
\multicolumn{9}{l}{$\star$ Age homogenization is measured by Shannon entropy over the distribution of selected patients across age brackets ($\uparrow$ entropy = $\downarrow$ homogenization).} \\
\addlinespace
\end{tabular} 
\caption{Evaluating recovered allocations from the Rashomon set. Avg over 10 train-test splits x 25 samples of $n=1000$ patients.}
\label{tab:results_main}
\end{table*}

\subsection{Evaluating Promises of the Rashomon Set}
For each Rashomon set sampling method, we compare its allocations to the full set of equal-utility allocations, revealing that the recovered allocations fall short of what the full space of equal-utility allocations offers.

\subsubsection*{\textbf{The least discriminatory allocation from the Rashomon set is far from the least discriminatory allocation with equal-utility that could be achieved.}} In real-world healthcare allocations, \citet{obermeyer2019dissecting} highlight the fact that selected Black patients are considerably sicker than selected White patients in terms of how many active chronic illnesses they have. This indicates discrimination based on the threshold test, which evaluates if the \textit{ratio of active chronic illnesses} between selected Black and White patients is greater than 1. In addition to its unfairness, this allocation is concerning from a prioritarian point of view: there are many Black patients who were not selected who are sicker than the White patients who were selected.  Across recovered allocations from the Rashomon set, we consider the least discriminatory allocation to have the smallest ratio based on the threshold test. Figure~\ref{fig:results_lda} shows that the smallest ratio recovered by all 4 methods is around 1.2, indicating that this ``least discriminatory allocation'' is still discriminatory. In contrast, the space of all equal-utility allocations contains many allocations with threshold test ratios at or less than 1. We specifically compare to an ideal equal-utility allocation that first selects $k'$ qualified patients in descending order by number of chronic illnesses, then selects $k-k'$ unqualified patients in the same fashion, alternating between Black and White patients when there are ties. As Figure~\ref{fig:results_lda} shows, this results in a least\footnote{We consider this the ``least discriminatory'' equal-utility allocation because it adheres to the normative goal of selecting patients based on how sick they are. Technically, an equal-utility allocation with a lower ratio could be obtained by adversarially choosing the least-ill Black patients and the most-ill White patients.} discriminatory equal-utility allocation with a threshold test ratio around 0.9.

\subsubsection*{\textbf{Even if decision-makers use different models in the Rashomon set, outcome homogenization remains high and some qualified individuals are still systemically rejected.}} We compare the pairwise consistency of individual outcomes between recovered allocations from the Rashomon set and all equal-utility allocations. \textit{Pairwise consistency} represents the probability an individual would receive the same outcome under two random allocations from the set under consideration~\citep{cooper2024arbitrariness}. As Table~\ref{tab:results_main} shows, the pairwise consistency is around 95\% for recovered allocations from the Rashomon set. This is much higher than the pairwise consistency across all equal-utility allocations, which is around 75\%. If decision-makers exclusively choose models from the Rashomon set, the high pairwise consistency implies that many individuals may be systemically rejected, even if they are qualified. Figure~\ref{fig:results_oh} analyzes how many qualified individuals receive multiple outcomes across recovered allocations from the Rashomon set. Depending on the method, between 10\% to 30\% of qualified individuals are never selected in any recovered allocation.

\subsubsection*{\textbf{The ensemble allocation from the Rashomon set is unrepresentative of all equal-utility allocations.}} Consider the scenario where a fairness-oriented decision-maker uses the Rashomon set to average predictions and then selects patients with the top $k$ highest predictions from this ensemble. We show that the selected individuals under this allocation have different features from the selected individuals under a randomly-drawn equal-utility allocation. We focus on a specific feature -- the age of selected patients -- because of its potential to be homogenized in healthcare allocations. A stark example is the UK's liver transplant matching algorithm, which systematically denied transplants to younger patients regardless of medical urgency~\citep{liverAgeDiscrimination}. Figure~\ref{fig:results_ensemble} shows a similar result for the ensemble allocation from the Rashomon set: the age of selected patients is skewed older. While older patients are expected to be sicker and more qualified, Figure~\ref{fig:results_ensemble} also shows the age distribution of selected patients under a randomly-drawn equal-utility allocation, which chooses significantly more younger patients. Table~\ref{tab:results_main} measures this phenomenon using \textit{Shannon entropy} over the distribution of selected patients across age brackets. The recovered allocations from the Rashomon set have a lower entropy, indicating more homogenization in the age of selected patients.

\section{Discussion}
%Why are the recovered allocations so small compared to the full space of equal-utility allocations?

Our results show that conventional methods of using the Rashomon set do not fulfill its promises in resource-constrained allocations. Since we have established that model multiplicity is useful primarily insofar as it yields allocation multiplicity, how can we find more equal-utility allocations? We discuss three approaches below as avenues for future research.

\subsubsection*{\textbf{Improving Sampling from the Rashomon Set}}

Allocation multiplicity should be an evaluation target for empirical methods that aim to find Rashomon models, especially given that a growing area of work involves developing and improving such methods~\citep{rudin2024position, black2022multiplicity}. As Table~\ref{tab:perf} shows, all the methods that we explore find many different models in the Rashomon set. However, some methods perform better than others in terms of how many unique allocations they recover under a top $k$ mapping. We find that using different subsets of features and training data recovers more unique allocations than shuffling data order or perturbing model weights. But given that the former methods involve re-training different models from scratch, there are computational limitations to their use in practice. Even when many unique allocations are recovered, we further show that they fall short of what the full-space of equal-utility allocations promises. We hope the perspective of allocation multiplicity inspires future work to develop methods of finding \textit{diverse} equally-good models that deliver on these promises. 

\subsubsection*{\textbf{Mitigating Structural Biases and Underspecification}}

\begin{figure}[t!]
\centering
\includegraphics[width=.9\linewidth]{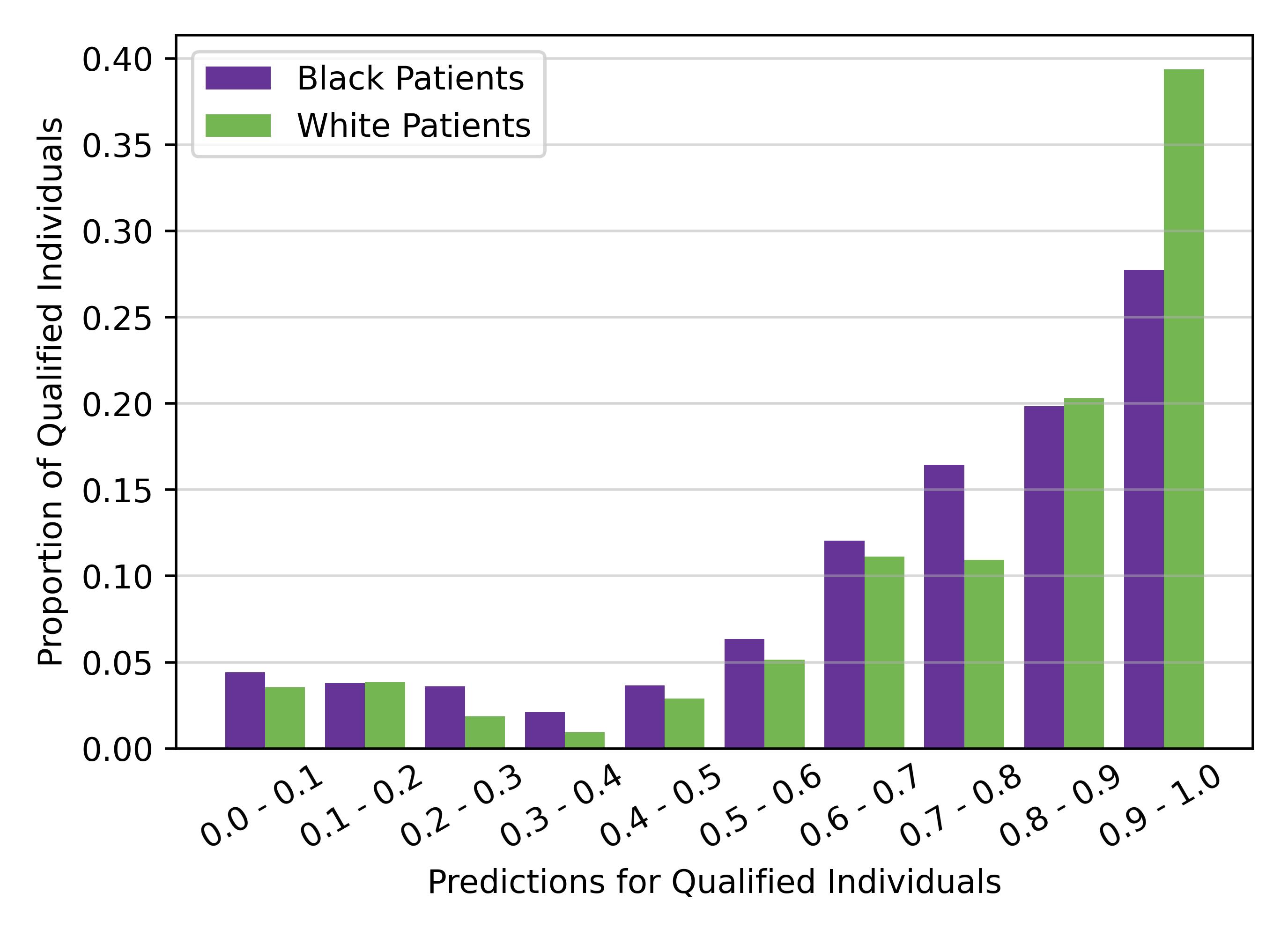}  
\caption{Among equally-sick patients, Black patients have systematically lower predictions than White patients across all equally-good models.}
\Description{Among equally-sick patients, Black patients have systematically lower predictions than White patients across all equally-good models.}
\label{fig:risk_scores}
\end{figure}

Addressing structural biases and underspecification in how risk scores are generated may lead to equal-utility allocations that status-quo models fail to reflect. Even if we sample extensively from the Rashomon set, the features used to model risk may lack the richness needed to distinguish between high-risk and low-risk individuals, particularly within certain subgroups. For example, in healthcare allocation, models trained on datasets that under-represent minority populations may fail to account for systemic barriers, such as limited access to preventive care or higher exposure to environmental risks~\citep{obermeyer2019dissecting}. As a result, these individuals are often misclassified as high-risk, even when they might benefit greatly from treatment (c.f. 
Figure~\ref{fig:risk_scores}). 
This deficiency is compounded by incomplete or biased training data that limits the ability of models to learn meaningful patterns for marginalized groups. Consequently, most models in the Rashomon set may converge on overly simplistic or biased risk estimates, systematically excluding certain qualified individuals and narrowing the recovered allocation space.

\subsubsection*{\textbf{Non-Deterministic Mappings Between Predictions and Decisions}}

Our case study focused on the limitations of allocations recovered by deterministically selecting individuals with the lowest risk. However, there are many stochastic procedures to map predictions to decisions~\citep{jain2024icml}. Decision-makers could use a weighted lottery based on predicted risk scores, or randomize decisions for individuals near the top $k$ threshold. Table~\ref{tab:results_main} shows that decision-boundary randomization outperforms all of the Rashomon set methods for each of the fairness metrics. In general, stochastic mappings may better explore the space of allocations than trying to find more models in the Rashomon set, which may not yield diverse allocations under the deterministic paradigm. While randomization involves trade-offs with \textit{expected} utility, there may be little or no trade-off with \textit{realized} utility, especially given the exponential number of equal-utility allocations that exist. In our case study, Table~\ref{tab:perf} shows that this trade-off is only around 1\% for decision-boundary randomization, despite yielding much larger fairness benefits. We hope that the knowledge of allocation multiplicity inspires a paradigm shift in how decision-makers interpret and act on model predictions, emphasizing flexibility and equity over rigid  risk-based thresholds.

\section*{Acknowledgements}
We are very grateful to Prakhar Ganesh, Jamelle Watson-Daniels, and Flavio Calmon for their feedback and comments on this paper. S.J. and A.W. are generously supported by the Simons Collaboration on The Theory of Algorithmic Fairness.

\bibliographystyle{ACM-Reference-Format}
\bibliography{refs}


\begin{thebibliography}{48}

%%% the Web address.

\ifx \showCODEN    \undefined \def \showCODEN     #1{\unskip}     \fi
\ifx \showDOI      \undefined \def \showDOI       #1{#1}\fi
\ifx \showISBNx    \undefined \def \showISBNx     #1{\unskip}     \fi
\ifx \showISBNxiii \undefined \def \showISBNxiii  #1{\unskip}     \fi
\ifx \showISSN     \undefined \def \showISSN      #1{\unskip}     \fi
\ifx \showLCCN     \undefined \def \showLCCN      #1{\unskip}     \fi
\ifx \shownote     \undefined \def \shownote      #1{#1}          \fi
\ifx \showarticletitle \undefined \def \showarticletitle #1{#1}   \fi
\ifx \showURL      \undefined \def \showURL       {\relax}        \fi
% invisible to TeX
\providecommand\bibfield[2]{#2}
\providecommand\bibinfo[2]{#2}
\providecommand\natexlab[1]{#1}
\providecommand\showeprint[2][]{arXiv:#2}

\bibitem[Antweiler(2018)]%
        {sigmoid_logit}
\bibfield{author}{\bibinfo{person}{Werner Antweiler}.} \bibinfo{year}{2018}\natexlab{}.
\newblock \bibinfo{title}{A sigmoid-logit probability function for the (0,1) domain}.
\newblock
\urldef\tempurl%
\url{https://wernerantweiler.ca/blog.php?item=2018-11-03}
\showURL{%
\tempurl}


\bibitem[Arneson(2022)]%
        {Arneson2022}
\bibfield{author}{\bibinfo{person}{Richard~J. Arneson}.} \bibinfo{year}{2022}\natexlab{}.
\newblock \bibinfo{booktitle}{\emph{Prioritarianism}}.
\newblock \bibinfo{publisher}{Cambridge University Press}.
\newblock
\showISBNx{9781108730693}
\urldef\tempurl%
\url{https://doi.org/10.1017/9781108582865}
\showDOI{\tempurl}


\bibitem[Auerbach et~al\mbox{.}(2024)]%
        {auerbach2024testing}
\bibfield{author}{\bibinfo{person}{Eric Auerbach}, \bibinfo{person}{Annie Liang}, \bibinfo{person}{Max Tabord-Meehan}, {and} \bibinfo{person}{Kyohei Okumura}.} \bibinfo{year}{2024}\natexlab{}.
\newblock \showarticletitle{Testing the Fairness-Improvability of Algorithms}.
\newblock \bibinfo{journal}{\emph{arXiv preprint arXiv:2405.04816}} (\bibinfo{year}{2024}).
\newblock


\bibitem[Barocas et~al\mbox{.}(2023)]%
        {barocas2023fairness}
\bibfield{author}{\bibinfo{person}{Solon Barocas}, \bibinfo{person}{Moritz Hardt}, {and} \bibinfo{person}{Arvind Narayanan}.} \bibinfo{year}{2023}\natexlab{}.
\newblock \bibinfo{booktitle}{\emph{Fairness and machine learning: Limitations and opportunities}}.
\newblock \bibinfo{publisher}{MIT press}.
\newblock


\bibitem[Black et~al\mbox{.}(2024a)]%
        {black2024d}
\bibfield{author}{\bibinfo{person}{Emily Black}, \bibinfo{person}{Talia Gillis}, {and} \bibinfo{person}{Zara~Yasmine Hall}.} \bibinfo{year}{2024}\natexlab{a}.
\newblock \showarticletitle{D-hacking}. In \bibinfo{booktitle}{\emph{The 2024 ACM Conference on Fairness, Accountability, and Transparency}}. \bibinfo{pages}{602--615}.
\newblock


\bibitem[Black et~al\mbox{.}(2024b)]%
        {black2024less}
\bibfield{author}{\bibinfo{person}{Emily Black}, \bibinfo{person}{John~Logan Koepke}, \bibinfo{person}{Pauline Kim}, \bibinfo{person}{Solon Barocas}, {and} \bibinfo{person}{Mingwei Hsu}.} \bibinfo{year}{2024}\natexlab{b}.
\newblock \showarticletitle{Less discriminatory algorithms}.
\newblock \bibinfo{journal}{\emph{Georgetown Law Journal}} (\bibinfo{year}{2024}).
\newblock
\urldef\tempurl%
\url{https://papers.ssrn.com/sol3/papers.cfm?abstract_id=4590481}
\showURL{%
\tempurl}


\bibitem[Black and Leino(2021)]%
        {black2021selective}
\bibfield{author}{\bibinfo{person}{Emily Black} {and} \bibinfo{person}{Klas Leino}.} \bibinfo{year}{2021}\natexlab{}.
\newblock \showarticletitle{Selective Ensembles for Consistent Predictions}. In \bibinfo{booktitle}{\emph{Ninth International Conference on Learning Representations}}.
\newblock


\bibitem[Black et~al\mbox{.}(2023)]%
        {black2023pipeline}
\bibfield{author}{\bibinfo{person}{Emily Black}, \bibinfo{person}{Rakshit Naidu}, \bibinfo{person}{Rayid Ghani}, \bibinfo{person}{Kit Rodolfa}, \bibinfo{person}{Daniel Ho}, {and} \bibinfo{person}{Hoda Heidari}.} \bibinfo{year}{2023}\natexlab{}.
\newblock \showarticletitle{Toward Operationalizing Pipeline-aware ML Fairness: A Research Agenda for Developing Practical Guidelines and Tools}. In \bibinfo{booktitle}{\emph{Proceedings of the 3rd ACM Conference on Equity and Access in Algorithms, Mechanisms, and Optimization}}. \bibinfo{pages}{1--11}.
\newblock


\bibitem[Black et~al\mbox{.}(2022)]%
        {black2022multiplicity}
\bibfield{author}{\bibinfo{person}{Emily Black}, \bibinfo{person}{Manish Raghavan}, {and} \bibinfo{person}{Solon Barocas}.} \bibinfo{year}{2022}\natexlab{}.
\newblock \showarticletitle{Model Multiplicity: Opportunities, Concerns, and Solutions}. In \bibinfo{booktitle}{\emph{Proceedings of the 2022 ACM Conference on Fairness, Accountability, and Transparency}} (Seoul, Republic of Korea) \emph{(\bibinfo{series}{FAccT '22})}. \bibinfo{publisher}{Association for Computing Machinery}, \bibinfo{address}{New York, NY, USA}, \bibinfo{pages}{850–863}.
\newblock
\showISBNx{9781450393522}
\urldef\tempurl%
\url{https://doi.org/10.1145/3531146.3533149}
\showDOI{\tempurl}


\bibitem[Bommasani et~al\mbox{.}(2022)]%
        {bommasani2022outcome}
\bibfield{author}{\bibinfo{person}{Rishi Bommasani}, \bibinfo{person}{Kathleen~A Creel}, \bibinfo{person}{Ananya Kumar}, \bibinfo{person}{Dan Jurafsky}, {and} \bibinfo{person}{Percy~S Liang}.} \bibinfo{year}{2022}\natexlab{}.
\newblock \showarticletitle{Picking on the same person: Does algorithmic monoculture lead to outcome homogenization?}
\newblock \bibinfo{journal}{\emph{Advances in Neural Information Processing Systems}}  \bibinfo{volume}{35} (\bibinfo{year}{2022}), \bibinfo{pages}{3663--3678}.
\newblock


\bibitem[Breiman(2001)]%
        {breiman2001statistical}
\bibfield{author}{\bibinfo{person}{Leo Breiman}.} \bibinfo{year}{2001}\natexlab{}.
\newblock \showarticletitle{Statistical modeling: The two cultures}.
\newblock \bibinfo{journal}{\emph{Statistical science}} \bibinfo{volume}{16}, \bibinfo{number}{3} (\bibinfo{year}{2001}), \bibinfo{pages}{199--231}.
\newblock


\bibitem[Broome(1990)]%
        {broome1990fairness}
\bibfield{author}{\bibinfo{person}{John Broome}.} \bibinfo{year}{1990}\natexlab{}.
\newblock \showarticletitle{Fairness}. In \bibinfo{booktitle}{\emph{Proceedings of the Aristotelian Society}}, Vol.~\bibinfo{volume}{91}. \bibinfo{pages}{87--101}.
\newblock


\bibitem[Cooper et~al\mbox{.}(2024)]%
        {cooper2024arbitrariness}
\bibfield{author}{\bibinfo{person}{A~Feder Cooper}, \bibinfo{person}{Katherine Lee}, \bibinfo{person}{Madiha~Zahrah Choksi}, \bibinfo{person}{Solon Barocas}, \bibinfo{person}{Christopher De~Sa}, \bibinfo{person}{James Grimmelmann}, \bibinfo{person}{Jon Kleinberg}, \bibinfo{person}{Siddhartha Sen}, {and} \bibinfo{person}{Baobao Zhang}.} \bibinfo{year}{2024}\natexlab{}.
\newblock \showarticletitle{Arbitrariness and social prediction: The confounding role of variance in fair classification}. In \bibinfo{booktitle}{\emph{Proceedings of the AAAI Conference on Artificial Intelligence}}, Vol.~\bibinfo{volume}{38}. \bibinfo{pages}{22004--22012}.
\newblock


\bibitem[Coston et~al\mbox{.}(2021)]%
        {coston2021characterizing}
\bibfield{author}{\bibinfo{person}{Amanda Coston}, \bibinfo{person}{Ashesh Rambachan}, {and} \bibinfo{person}{Alexandra Chouldechova}.} \bibinfo{year}{2021}\natexlab{}.
\newblock \showarticletitle{Characterizing fairness over the set of good models under selective labels}. In \bibinfo{booktitle}{\emph{International Conference on Machine Learning}}. PMLR, \bibinfo{pages}{2144--2155}.
\newblock


\bibitem[Creel and Hellman(2022)]%
        {creel_hellman_2022}
\bibfield{author}{\bibinfo{person}{Kathleen Creel} {and} \bibinfo{person}{Deborah Hellman}.} \bibinfo{year}{2022}\natexlab{}.
\newblock \showarticletitle{The Algorithmic Leviathan: Arbitrariness, Fairness, and Opportunity in Algorithmic Decision-Making Systems}.
\newblock \bibinfo{journal}{\emph{Canadian Journal of Philosophy}} \bibinfo{volume}{52}, \bibinfo{number}{1} (\bibinfo{year}{2022}), \bibinfo{pages}{26–43}.
\newblock
\urldef\tempurl%
\url{https://doi.org/10.1017/can.2022.3}
\showDOI{\tempurl}


\bibitem[D'Amour et~al\mbox{.}(2022)]%
        {d2022underspecification}
\bibfield{author}{\bibinfo{person}{Alexander D'Amour}, \bibinfo{person}{Katherine Heller}, \bibinfo{person}{Dan Moldovan}, \bibinfo{person}{Ben Adlam}, \bibinfo{person}{Babak Alipanahi}, \bibinfo{person}{Alex Beutel}, \bibinfo{person}{Christina Chen}, \bibinfo{person}{Jonathan Deaton}, \bibinfo{person}{Jacob Eisenstein}, \bibinfo{person}{Matthew~D Hoffman}, {et~al\mbox{.}}} \bibinfo{year}{2022}\natexlab{}.
\newblock \showarticletitle{Underspecification presents challenges for credibility in modern machine learning}.
\newblock \bibinfo{journal}{\emph{Journal of Machine Learning Research}} \bibinfo{volume}{23}, \bibinfo{number}{226} (\bibinfo{year}{2022}), \bibinfo{pages}{1--61}.
\newblock


\bibitem[Dawid(2017)]%
        {dawid2017individual}
\bibfield{author}{\bibinfo{person}{Philip Dawid}.} \bibinfo{year}{2017}\natexlab{}.
\newblock \showarticletitle{On individual risk}.
\newblock \bibinfo{journal}{\emph{Synthese}} \bibinfo{volume}{194}, \bibinfo{number}{9} (\bibinfo{year}{2017}), \bibinfo{pages}{3445--3474}.
\newblock


\bibitem[Du et~al\mbox{.}(2024)]%
        {du2024reconciling}
\bibfield{author}{\bibinfo{person}{Ally~Yalei Du}, \bibinfo{person}{Dung~Daniel Ngo}, {and} \bibinfo{person}{Zhiwei~Steven Wu}.} \bibinfo{year}{2024}\natexlab{}.
\newblock \showarticletitle{Reconciling Model Multiplicity for Downstream Decision Making}.
\newblock \bibinfo{journal}{\emph{arXiv preprint arXiv:2405.19667}} (\bibinfo{year}{2024}).
\newblock


\bibitem[Dwork et~al\mbox{.}(2021)]%
        {dwork2021outcome}
\bibfield{author}{\bibinfo{person}{Cynthia Dwork}, \bibinfo{person}{Michael~P Kim}, \bibinfo{person}{Omer Reingold}, \bibinfo{person}{Guy~N Rothblum}, {and} \bibinfo{person}{Gal Yona}.} \bibinfo{year}{2021}\natexlab{}.
\newblock \showarticletitle{Outcome indistinguishability}. In \bibinfo{booktitle}{\emph{Proceedings of the 53rd Annual ACM SIGACT Symposium on Theory of Computing}}. \bibinfo{pages}{1095--1108}.
\newblock


\bibitem[Fisher et~al\mbox{.}(2019)]%
        {fisher2019all}
\bibfield{author}{\bibinfo{person}{Aaron Fisher}, \bibinfo{person}{Cynthia Rudin}, {and} \bibinfo{person}{Francesca Dominici}.} \bibinfo{year}{2019}\natexlab{}.
\newblock \showarticletitle{All models are wrong, but many are useful: Learning a variable's importance by studying an entire class of prediction models simultaneously}.
\newblock \bibinfo{journal}{\emph{Journal of Machine Learning Research}} \bibinfo{volume}{20}, \bibinfo{number}{177} (\bibinfo{year}{2019}), \bibinfo{pages}{1--81}.
\newblock


\bibitem[Ganesh(2024)]%
        {ganesh2024empirical}
\bibfield{author}{\bibinfo{person}{Prakhar Ganesh}.} \bibinfo{year}{2024}\natexlab{}.
\newblock \showarticletitle{An Empirical Investigation into Benchmarking Model Multiplicity for Trustworthy Machine Learning: A Case Study on Image Classification}. In \bibinfo{booktitle}{\emph{Proceedings of the IEEE/CVF Winter Conference on Applications of Computer Vision}}. \bibinfo{pages}{4488--4497}.
\newblock


\bibitem[Ganesh et~al\mbox{.}(2023)]%
        {ganesh2023impact}
\bibfield{author}{\bibinfo{person}{Prakhar Ganesh}, \bibinfo{person}{Hongyan Chang}, \bibinfo{person}{Martin Strobel}, {and} \bibinfo{person}{Reza Shokri}.} \bibinfo{year}{2023}\natexlab{}.
\newblock \showarticletitle{On the impact of machine learning randomness on group fairness}. In \bibinfo{booktitle}{\emph{Proceedings of the 2023 ACM Conference on Fairness, Accountability, and Transparency}}. \bibinfo{pages}{1789--1800}.
\newblock


\bibitem[Gillis et~al\mbox{.}(2024)]%
        {gillis2024operationalizing}
\bibfield{author}{\bibinfo{person}{Talia~B Gillis}, \bibinfo{person}{Vitaly Meursault}, {and} \bibinfo{person}{Berk Ustun}.} \bibinfo{year}{2024}\natexlab{}.
\newblock \showarticletitle{Operationalizing the Search for Less Discriminatory Alternatives in Fair Lending}. In \bibinfo{booktitle}{\emph{The 2024 ACM Conference on Fairness, Accountability, and Transparency}}. \bibinfo{pages}{377--387}.
\newblock


\bibitem[H{\'e}bert-Johnson et~al\mbox{.}(2018)]%
        {hebert2018multicalibration}
\bibfield{author}{\bibinfo{person}{Ursula H{\'e}bert-Johnson}, \bibinfo{person}{Michael Kim}, \bibinfo{person}{Omer Reingold}, {and} \bibinfo{person}{Guy Rothblum}.} \bibinfo{year}{2018}\natexlab{}.
\newblock \showarticletitle{Multicalibration: Calibration for the (computationally-identifiable) masses}. In \bibinfo{booktitle}{\emph{International Conference on Machine Learning}}. PMLR, \bibinfo{pages}{1939--1948}.
\newblock


\bibitem[Hsu and Calmon(2022)]%
        {hsu2022rashomon}
\bibfield{author}{\bibinfo{person}{Hsiang Hsu} {and} \bibinfo{person}{Flavio Calmon}.} \bibinfo{year}{2022}\natexlab{}.
\newblock \showarticletitle{Rashomon capacity: A metric for predictive multiplicity in classification}.
\newblock \bibinfo{journal}{\emph{Advances in Neural Information Processing Systems}}  \bibinfo{volume}{35} (\bibinfo{year}{2022}), \bibinfo{pages}{28988--29000}.
\newblock
\urldef\tempurl%
\url{https://proceedings.neurips.cc/paper_files/paper/2022/file/ba4caa85ecdcafbf9102ab8ec384182d-Paper-Conference.pdf}
\showURL{%
\tempurl}


\bibitem[Hsu et~al\mbox{.}(2024)]%
        {hsu2024dropout}
\bibfield{author}{\bibinfo{person}{Hsiang Hsu}, \bibinfo{person}{Guihong Li}, \bibinfo{person}{Shaohan Hu}, {et~al\mbox{.}}} \bibinfo{year}{2024}\natexlab{}.
\newblock \showarticletitle{Dropout-Based Rashomon Set Exploration for Efficient Predictive Multiplicity Estimation}.
\newblock \bibinfo{journal}{\emph{arXiv preprint arXiv:2402.00728}} (\bibinfo{year}{2024}).
\newblock


\bibitem[Jain et~al\mbox{.}(2024a)]%
        {jain2024icml}
\bibfield{author}{\bibinfo{person}{Shomik Jain}, \bibinfo{person}{Kathleen Creel}, {and} \bibinfo{person}{Ashia Wilson}.} \bibinfo{year}{2024}\natexlab{a}.
\newblock \showarticletitle{Position: Scarce Resource Allocations That Rely On Machine Learning Should Be Randomized}. In \bibinfo{booktitle}{\emph{Forty-first International Conference on Machine Learning}}.
\newblock


\bibitem[Jain et~al\mbox{.}(2024b)]%
        {jain2024facct}
\bibfield{author}{\bibinfo{person}{Shomik Jain}, \bibinfo{person}{Vinith Suriyakumar}, \bibinfo{person}{Kathleen Creel}, {and} \bibinfo{person}{Ashia Wilson}.} \bibinfo{year}{2024}\natexlab{b}.
\newblock \showarticletitle{Algorithmic Pluralism: A Structural Approach To Equal Opportunity}. In \bibinfo{booktitle}{\emph{The 2024 ACM Conference on Fairness, Accountability, and Transparency}}. \bibinfo{pages}{197--206}.
\newblock


\bibitem[Jiang et~al\mbox{.}(2023)]%
        {jiang2023recourse}
\bibfield{author}{\bibinfo{person}{Junqi Jiang}, \bibinfo{person}{Antonio Rago}, \bibinfo{person}{Francesco Leofante}, {and} \bibinfo{person}{Francesca Toni}.} \bibinfo{year}{2023}\natexlab{}.
\newblock \showarticletitle{Recourse under model multiplicity via argumentative ensembling}.
\newblock \bibinfo{journal}{\emph{arXiv preprint arXiv:2312.15097}} (\bibinfo{year}{2023}).
\newblock


\bibitem[Kleinberg and Raghavan(2021)]%
        {kleinberg2021algorithmic}
\bibfield{author}{\bibinfo{person}{Jon Kleinberg} {and} \bibinfo{person}{Manish Raghavan}.} \bibinfo{year}{2021}\natexlab{}.
\newblock \showarticletitle{Algorithmic monoculture and social welfare}.
\newblock \bibinfo{journal}{\emph{Proceedings of the National Academy of Sciences}} \bibinfo{volume}{118}, \bibinfo{number}{22} (\bibinfo{year}{2021}), \bibinfo{pages}{e2018340118}.
\newblock


\bibitem[Laufer et~al\mbox{.}(2024)]%
        {laufer2024fundamental}
\bibfield{author}{\bibinfo{person}{Benjamin Laufer}, \bibinfo{person}{Manish Raghavan}, {and} \bibinfo{person}{Solon Barocas}.} \bibinfo{year}{2024}\natexlab{}.
\newblock \showarticletitle{Fundamental Limits in the Search for Less Discriminatory Algorithms--and How to Avoid Them}.
\newblock \bibinfo{journal}{\emph{arXiv preprint arXiv:2412.18138}} (\bibinfo{year}{2024}).
\newblock


\bibitem[Li et~al\mbox{.}(2024)]%
        {li2024practical}
\bibfield{author}{\bibinfo{person}{Sichao Li}, \bibinfo{person}{Amanda~S Barnard}, {and} \bibinfo{person}{Quanling Deng}.} \bibinfo{year}{2024}\natexlab{}.
\newblock \showarticletitle{Practical attribution guidance for rashomon sets}.
\newblock \bibinfo{journal}{\emph{arXiv preprint arXiv:2407.18482}} (\bibinfo{year}{2024}).
\newblock


\bibitem[Liu et~al\mbox{.}(2022)]%
        {liu2022fasterrisk}
\bibfield{author}{\bibinfo{person}{Jiachang Liu}, \bibinfo{person}{Chudi Zhong}, \bibinfo{person}{Boxuan Li}, \bibinfo{person}{Margo Seltzer}, {and} \bibinfo{person}{Cynthia Rudin}.} \bibinfo{year}{2022}\natexlab{}.
\newblock \showarticletitle{FasterRisk: Fast and Accurate Interpretable Risk Scores}. In \bibinfo{booktitle}{\emph{Advances in Neural Information Processing Systems}}, \bibfield{editor}{\bibinfo{person}{Alice~H. Oh}, \bibinfo{person}{Alekh Agarwal}, \bibinfo{person}{Danielle Belgrave}, {and} \bibinfo{person}{Kyunghyun Cho}} (Eds.).
\newblock
\urldef\tempurl%
\url{https://openreview.net/forum?id=xTYL1J6Xt-z}
\showURL{%
\tempurl}


\bibitem[Long et~al\mbox{.}(2024)]%
        {long2024individual}
\bibfield{author}{\bibinfo{person}{Carol Long}, \bibinfo{person}{Hsiang Hsu}, \bibinfo{person}{Wael Alghamdi}, {and} \bibinfo{person}{Flavio Calmon}.} \bibinfo{year}{2024}\natexlab{}.
\newblock \showarticletitle{Individual arbitrariness and group fairness}.
\newblock \bibinfo{journal}{\emph{Advances in Neural Information Processing Systems}}  \bibinfo{volume}{36} (\bibinfo{year}{2024}).
\newblock


\bibitem[Marx et~al\mbox{.}(2020)]%
        {marx_predictive_2020}
\bibfield{author}{\bibinfo{person}{Charles Marx}, \bibinfo{person}{Flavio Calmon}, {and} \bibinfo{person}{Berk Ustun}.} \bibinfo{year}{2020}\natexlab{}.
\newblock \showarticletitle{Predictive Multiplicity in Classification}. In \bibinfo{booktitle}{\emph{Proceedings of the 37th International Conference on Machine Learning}} \emph{(\bibinfo{series}{Proceedings of Machine Learning Research}, Vol.~\bibinfo{volume}{119})}, \bibfield{editor}{\bibinfo{person}{Hal~Daumé III} {and} \bibinfo{person}{Aarti Singh}} (Eds.). \bibinfo{publisher}{PMLR}, \bibinfo{pages}{6765--6774}.
\newblock
\urldef\tempurl%
\url{https://proceedings.mlr.press/v119/marx20a.html}
\showURL{%
\tempurl}


\bibitem[Meyer et~al\mbox{.}(2023)]%
        {meyer2023dataset}
\bibfield{author}{\bibinfo{person}{Anna~P Meyer}, \bibinfo{person}{Aws Albarghouthi}, {and} \bibinfo{person}{Loris D'Antoni}.} \bibinfo{year}{2023}\natexlab{}.
\newblock \showarticletitle{The dataset multiplicity problem: How unreliable data impacts predictions}. In \bibinfo{booktitle}{\emph{Proceedings of the 2023 ACM Conference on Fairness, Accountability, and Transparency}}. \bibinfo{pages}{193--204}.
\newblock


\bibitem[Narayanan et~al\mbox{.}(2024)]%
        {liverAgeDiscrimination}
\bibfield{author}{\bibinfo{person}{Arvind Narayanan}, \bibinfo{person}{Angelina Wang}, \bibinfo{person}{Sayash Kapoor}, {and} \bibinfo{person}{Solon Barocas}.} \bibinfo{year}{2024}\natexlab{}.
\newblock \bibinfo{title}{Does the UK’s liver transplant matching algorithm systematically exclude younger patients?}
\newblock \bibinfo{howpublished}{\url{https://www.aisnakeoil.com/p/does-the-uks-liver-transplant-matching}}.
\newblock


\bibitem[Obermeyer et~al\mbox{.}(2019)]%
        {obermeyer2019dissecting}
\bibfield{author}{\bibinfo{person}{Ziad Obermeyer}, \bibinfo{person}{Brian Powers}, \bibinfo{person}{Christine Vogeli}, {and} \bibinfo{person}{Sendhil Mullainathan}.} \bibinfo{year}{2019}\natexlab{}.
\newblock \showarticletitle{Dissecting racial bias in an algorithm used to manage the health of populations}.
\newblock \bibinfo{journal}{\emph{Science}} \bibinfo{volume}{366}, \bibinfo{number}{6464} (\bibinfo{year}{2019}), \bibinfo{pages}{447--453}.
\newblock


\bibitem[Roth et~al\mbox{.}(2023)]%
        {roth2023reconciling}
\bibfield{author}{\bibinfo{person}{Aaron Roth}, \bibinfo{person}{Alexander Tolbert}, {and} \bibinfo{person}{Scott Weinstein}.} \bibinfo{year}{2023}\natexlab{}.
\newblock \showarticletitle{Reconciling Individual Probability Forecasts}. In \bibinfo{booktitle}{\emph{Proceedings of the 2023 ACM Conference on Fairness, Accountability, and Transparency}}. \bibinfo{pages}{101--110}.
\newblock


\bibitem[Rudin et~al\mbox{.}(2024)]%
        {rudin2024position}
\bibfield{author}{\bibinfo{person}{Cynthia Rudin}, \bibinfo{person}{Chudi Zhong}, \bibinfo{person}{Lesia Semenova}, \bibinfo{person}{Margo Seltzer}, \bibinfo{person}{Ronald Parr}, \bibinfo{person}{Jiachang Liu}, \bibinfo{person}{Srikar Katta}, \bibinfo{person}{Jon Donnelly}, \bibinfo{person}{Harry Chen}, {and} \bibinfo{person}{Zachery Boner}.} \bibinfo{year}{2024}\natexlab{}.
\newblock \showarticletitle{Position: Amazing Things Come From Having Many Good Models}. In \bibinfo{booktitle}{\emph{Forty-first International Conference on Machine Learning}}.
\newblock
\urldef\tempurl%
\url{https://openreview.net/forum?id=oFDFGd9Age}
\showURL{%
\tempurl}


\bibitem[Semenova et~al\mbox{.}(2023)]%
        {semenova2023path}
\bibfield{author}{\bibinfo{person}{Lesia Semenova}, \bibinfo{person}{Harry Chen}, \bibinfo{person}{Ronald Parr}, {and} \bibinfo{person}{Cynthia Rudin}.} \bibinfo{year}{2023}\natexlab{}.
\newblock \showarticletitle{A path to simpler models starts with noise}.
\newblock \bibinfo{journal}{\emph{Advances in neural information processing systems}}  \bibinfo{volume}{36} (\bibinfo{year}{2023}).
\newblock
\urldef\tempurl%
\url{https://proceedings.neurips.cc/paper_files/paper/2023/file/0a49935d2b3d3342ca08d6db0adcfa34-Paper-Conference.pdf}
\showURL{%
\tempurl}


\bibitem[Semenova et~al\mbox{.}(2022)]%
        {semenova2022existence}
\bibfield{author}{\bibinfo{person}{Lesia Semenova}, \bibinfo{person}{Cynthia Rudin}, {and} \bibinfo{person}{Ronald Parr}.} \bibinfo{year}{2022}\natexlab{}.
\newblock \showarticletitle{On the existence of simpler machine learning models}. In \bibinfo{booktitle}{\emph{Proceedings of the 2022 ACM Conference on Fairness, Accountability, and Transparency}}. \bibinfo{pages}{1827--1858}.
\newblock


\bibitem[Simson et~al\mbox{.}(2024)]%
        {simson2024one}
\bibfield{author}{\bibinfo{person}{Jan Simson}, \bibinfo{person}{Florian Pfisterer}, {and} \bibinfo{person}{Christoph Kern}.} \bibinfo{year}{2024}\natexlab{}.
\newblock \showarticletitle{One model many scores: Using multiverse analysis to prevent fairness hacking and evaluate the influence of model design decisions}. In \bibinfo{booktitle}{\emph{The 2024 ACM Conference on Fairness, Accountability, and Transparency}}. \bibinfo{pages}{1305--1320}.
\newblock


\bibitem[Toups et~al\mbox{.}(2023)]%
        {toups2023ecosystem}
\bibfield{author}{\bibinfo{person}{Connor Toups}, \bibinfo{person}{Rishi Bommasani}, \bibinfo{person}{Kathleen Creel}, \bibinfo{person}{Sarah Bana}, \bibinfo{person}{Dan Jurafsky}, {and} \bibinfo{person}{Percy~S Liang}.} \bibinfo{year}{2023}\natexlab{}.
\newblock \showarticletitle{Ecosystem-level analysis of deployed machine learning reveals homogeneous outcomes}.
\newblock \bibinfo{journal}{\emph{Advances in Neural Information Processing Systems}}  \bibinfo{volume}{36} (\bibinfo{year}{2023}).
\newblock


\bibitem[Watson-Daniels et~al\mbox{.}(2023a)]%
        {watson2023multi}
\bibfield{author}{\bibinfo{person}{Jamelle Watson-Daniels}, \bibinfo{person}{Solon Barocas}, \bibinfo{person}{Jake~M Hofman}, {and} \bibinfo{person}{Alexandra Chouldechova}.} \bibinfo{year}{2023}\natexlab{a}.
\newblock \showarticletitle{Multi-target multiplicity: Flexibility and fairness in target specification under resource constraints}. In \bibinfo{booktitle}{\emph{Proceedings of the 2023 ACM Conference on Fairness, Accountability, and Transparency}}. \bibinfo{pages}{297--311}.
\newblock
\urldef\tempurl%
\url{https://dl.acm.org/doi/pdf/10.1145/3593013.3593998}
\showURL{%
\tempurl}


\bibitem[Watson-Daniels et~al\mbox{.}(2023b)]%
        {watson2023predictive}
\bibfield{author}{\bibinfo{person}{Jamelle Watson-Daniels}, \bibinfo{person}{David~C Parkes}, {and} \bibinfo{person}{Berk Ustun}.} \bibinfo{year}{2023}\natexlab{b}.
\newblock \showarticletitle{Predictive multiplicity in probabilistic classification}. In \bibinfo{booktitle}{\emph{Proceedings of the AAAI Conference on Artificial Intelligence}}, Vol.~\bibinfo{volume}{37}. \bibinfo{pages}{10306--10314}.
\newblock


\bibitem[Xin et~al\mbox{.}(2022)]%
        {xin2022exploring}
\bibfield{author}{\bibinfo{person}{Rui Xin}, \bibinfo{person}{Chudi Zhong}, \bibinfo{person}{Zhi Chen}, \bibinfo{person}{Takuya Takagi}, \bibinfo{person}{Margo Seltzer}, {and} \bibinfo{person}{Cynthia Rudin}.} \bibinfo{year}{2022}\natexlab{}.
\newblock \showarticletitle{Exploring the Whole Rashomon Set of Sparse Decision Trees}. In \bibinfo{booktitle}{\emph{Advances in Neural Information Processing Systems}}, \bibfield{editor}{\bibinfo{person}{Alice~H. Oh}, \bibinfo{person}{Alekh Agarwal}, \bibinfo{person}{Danielle Belgrave}, {and} \bibinfo{person}{Kyunghyun Cho}} (Eds.).
\newblock
\urldef\tempurl%
\url{https://openreview.net/forum?id=WHqVVk3UHr}
\showURL{%
\tempurl}


\bibitem[Zhong et~al\mbox{.}(2023)]%
        {zhong2023exploring}
\bibfield{author}{\bibinfo{person}{Chudi Zhong}, \bibinfo{person}{Zhi Chen}, \bibinfo{person}{Jiachang Liu}, \bibinfo{person}{Margo Seltzer}, {and} \bibinfo{person}{Cynthia Rudin}.} \bibinfo{year}{2023}\natexlab{}.
\newblock \showarticletitle{Exploring and Interacting with the Set of Good Sparse Generalized Additive Models}. In \bibinfo{booktitle}{\emph{Thirty-seventh Conference on Neural Information Processing Systems}}.
\newblock
\urldef\tempurl%
\url{https://openreview.net/forum?id=CzAAbKOHQW}
\showURL{%
\tempurl}


\end{thebibliography}

\appendix

\clearpage

\section*{Appendix}

The Appendix includes the following supplementary tables and figures. 

\subsubsection*{\textbf{Table~\ref{tab:features}}} Selected features from the~\citet{obermeyer2019dissecting} dataset used to train models in all experiments. We use the same subset of features as~\citet{watson2023multi}. In particular, we do not use race as a feature.

\subsubsection*{\textbf{Table~\ref{tab:all_perf}}} Performance metrics for different methods of sampling from the Rashomon set. Expanded version of Table~\ref{tab:perf}, with additional results for $k/n=0.10$ and $n'/n=0.19$, as well as standard deviation results over experiment runs.

\begin{itemize}
    \item The different feature subsets method has a slightly lower allocation utility compared to other methods, despite having a the lowest validation loss on average.
    \item All methods have an average validation loss and deployment utility within 0.01 of each other, for all selection rates and qualification rates. 
\end{itemize}

\subsubsection*{\textbf{Table~\ref{tab:all_results}}} Evaluating recovered allocations from the Rashomon set. Expanded version of Table~\ref{tab:results_main}, with additional results for $k/n=0.10$ and $n'/n=0.19$, as well as standard deviation results over experiment runs. 

\begin{itemize}
    \item Generally, the promises of the Rashomon Set are least fulfilled when scarcity is highest. In particular, the yielded benefits from the Rashomon Set are farthest from the potential benefits under all equal-utility allocations for the qualification rate of $n'/n = 0.55$.
    \item For some selection rates and qualification rates, the threshold test ratio from the different feature subsets method is lower than the ratio reported for all equal-utility allocations. This is because the all equal-utility allocations ratio is computed based on the normative ideal of selecting patients based on how sick they are (in descending order by number of chronic illnesses). Technically, there exist equal-utility allocations with a lower ratio if the least-ill Black patients and most-ill White patients are selected, which the models found by the different feature subsets method may reflect. 

\end{itemize}

\subsubsection*{\textbf{Table~\ref{tab:outcome_homog}}} Additional metrics for outcome homogenization over recovered allocations from the Rashomon set. The table reports the proportion of qualified individuals that 1) are rejected in all recovered allocations, 2) receive multiple outcomes across recovered allocations, and 3) are accepted in all recovered allocations. 

\begin{itemize}
    \item The different feature subsets and bootstrapped training data methods have the lowest rates of systemic rejection and highest rates of qualified individuals receiving multiple outcomes across the Rashomon Set. 
    \item When there is scarcity ($k/n < n'/n$), the systemic rejection rate is high for all methods.
\end{itemize}

\subsubsection*{\textbf{Table~\ref{tab:all_random_results}}} Evaluating how stochastic mappings between predictions and decisions compare to top $k$ allocations from the Rashomon set. 

\begin{figure}[h!]
\centering
\includegraphics[width=\linewidth]{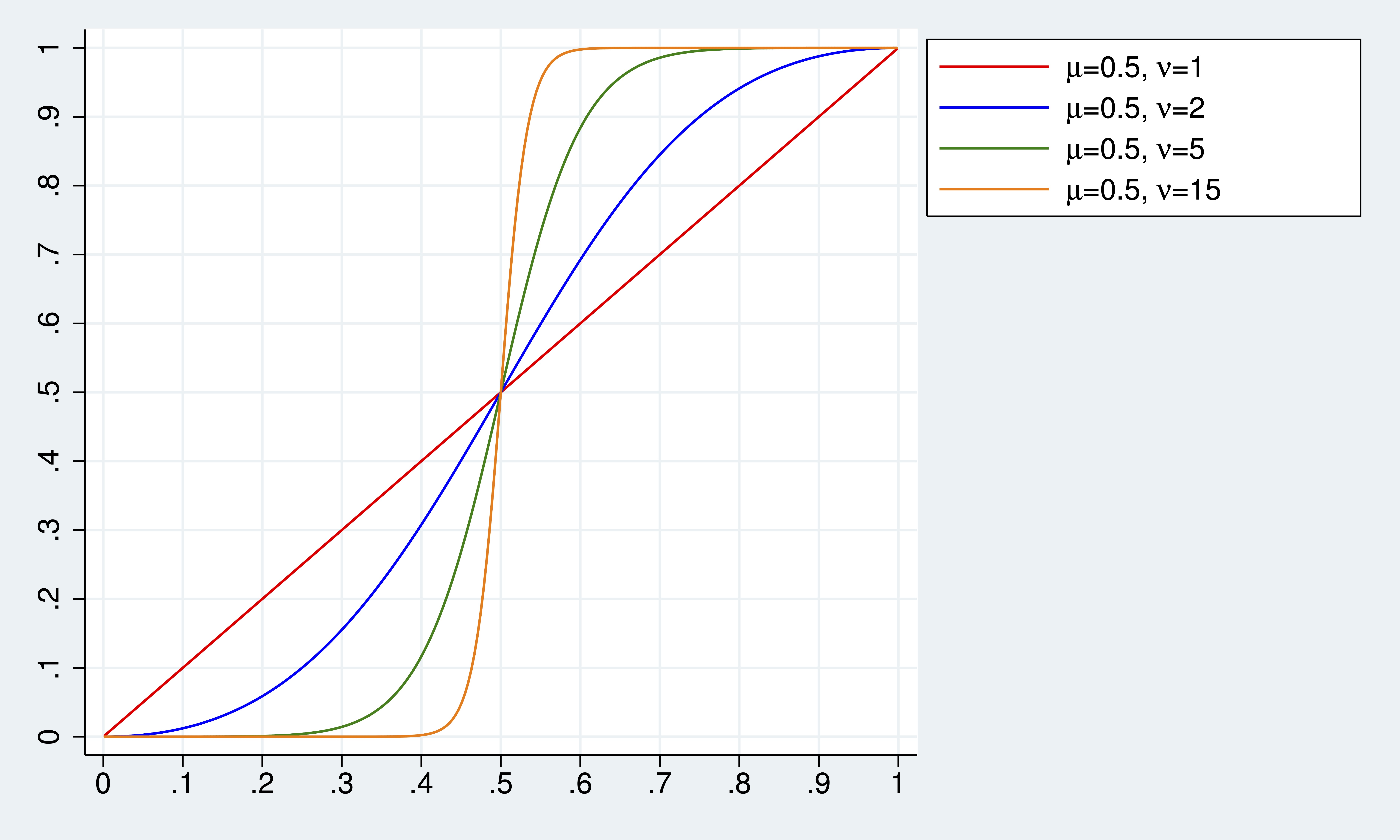}  
\caption{Sigmoid-logit function for varying $\mu$ and $v$ (Figure from \citet{sigmoid_logit})}
\label{fig:sigmoid_logit}
\end{figure}

\begin{itemize}
    \item Decision-boundary randomization refers to the partial B.F. lottery method from \citet{jain2024icml}. Specifically, $\tilde{k}$ resources are randomly allocated over $\tilde{n}$ people. First, $k-\tilde{k}$ resources are deterministically given to individuals with the highest predictions. Then, an iterative weighted selection is conducted for the remaining $\tilde{k}$ resources over the next $\tilde{n}$ highest predictions. We try two parameter choices: ($\tilde{k}=0.25k, \tilde{n}=0.50k$) and ($\tilde{k}=0.50k, \tilde{n}=1.0k$).
    \item Sigmoid-Logit randomization refers to a weighted lottery where the weights are determined using a sigmoid-logit transformation of predicted scores. The sigmoid-logit function~\cite{sigmoid_logit} is defined as $$f(x;\mu, v) = \left[1+\left(\frac{x \cdot (1-\mu)}{\mu \cdot (1-x)}\right)^{-v}\right]^{-1}$$ Figure~\ref{fig:sigmoid_logit} illustrates the function for different values of $\mu$ and $v$. We take $\mu = 1 - k/n$ and try two different values for $v$: $v=2$ and $v=5$. In a weighted lottery where the weights are based on predicted risk scores, the sigmoid-lottery function up-weights the probability that individuals in the top $k$ are selected and down-weights the probability that individuals in the bottom $n-k$ are selected. Figure~\ref{fig:sigmoid_logit} plots the sigmoid-logit function for different values of $v$.
 \end{itemize}

\begin{table*}[t!]
\small
\centering
\caption{Healthcare Data Dictionary (\citet{obermeyer2019dissecting})}
\begin{tabular}{l|l} 
\toprule
Feature & Description \\
\midrule
dem\_female & Indicator for female gender\\
dem\_age\_band\_18-24\_tm1 & Indicator for patient age between 18-24 \\
dem\_age\_band\_25-34\_tm1 & Indicator for patient age between 25-34 \\
dem\_age\_band\_35-44\_tm1 & Indicator for patient age between 35-44 \\
dem\_age\_band\_45-54\_tm1 & Indicator for patient age between 45-54 \\
dem\_age\_band\_55-64\_tm1 & Indicator for patient age between 55-64 \\
dem\_age\_band\_65-74\_tm1 & Indicator for patient age between 65-74 \\
dem\_age\_band\_75+\_tm1 & Indicator for patient age 75+ \\
hypertension\_elixhauser\_tm1 & Indicator for hypertension \\
cost\_dialysis\_tm1 & Total costs for dialysis, rounded to nearest 10 \\
cost\_emergency\_tm1 & Total costs for emergency, rounded to nearest 10 \\
cost\_home\_health\_tm1 & Total costs for home health, rounded to nearest 10 \\
cost\_ip\_medical\_tm1 & Total costs for inpatient medical, rounded to nearest 10 \\
cost\_ip\_surgical\_tm1 & Total costs for inpatient surgical, rounded to nearest 10 \\
cost\_laboratory\_tm1 & Total costs for laboratory, rounded to nearest 10 \\
cost\_op\_primary\_care\_tm1 & Total costs for outpatient primary care, rounded to nearest 10 \\
cost\_op\_specialists\_tm1 & Total costs for outpatient specialists, rounded to nearest 10 \\
cost\_op\_surgery\_tm1 & Total costs for outpatient surgery, rounded to nearest 10 \\
cost\_other\_tm1 & Total other costs, rounded to nearest 100 \\
cost\_pharmacy\_tm1 & Total costs for pharmacy, rounded to nearest 10 \\
cost\_physical\_therapy\_tm1 & Total costs for physical therapy, rounded to nearest 10 \\
cost\_radiology\_tm1 & Total costs for radiology, rounded to nearest 10 \\
gagne\_sum\_tm1 & Total number of active illnesses \\
\bottomrule
\end{tabular} 
\label{tab:features}
\end{table*}

\begin{table*}[b!]
\small
\centering
\caption{Performance Metrics For Different Methods of Sampling From the Rashomon Set (c.f. Table~\ref{tab:perf})}
\begin{tabular}{c|cc|cccc} 
\toprule
\multirow{2}{*}{Metric (Mean $\pm$ SD)} & \multirow{2}{*}{$k/n$} & \multirow{2}{*}{$n'/n$} & Different & Bootstrapped & Shuffled & Weight \\
& &  & Feature Subsets & Training Data & Data Order & Perturbation \\
\toprule
\addlinespace
\multirow{3}{*}{\shortstack[c]{Number of\\ Rashomon Models}}   & - & 0.19    & 1717 $\pm$ 3 & 995 $\pm$ 8 & 1000 $\pm$ 0 & 997 $\pm$ 5 \\
 & - & 0.32    & 1674 $\pm$ 3 & 997 $\pm$ 4 & 959 $\pm$ 25 & 999 $\pm$ 2 \\
 & - & 0.55    & 1386 $\pm$ 3 & 1000 $\pm$ 0 & 835 $\pm$ 62 & 1000 $\pm$ 0 \\
\addlinespace
\midrule
\addlinespace
\multirow{3}{*}{Validation Loss}   & - & 0.19    & 0.218 $\pm$ 0.006 & 0.223 $\pm$ 0.006 & 0.222 $\pm$ 0.006 & 0.227 $\pm$ 0.006 \\
 & - & 0.32    & 0.279 $\pm$ 0.006 & 0.284 $\pm$ 0.006 & 0.284 $\pm$ 0.006 & 0.290 $\pm$ 0.006 \\
 & - & 0.55    & 0.327 $\pm$ 0.006 & 0.328 $\pm$ 0.007 & 0.329 $\pm$ 0.006 & 0.335 $\pm$ 0.006 \\
\addlinespace
\midrule
\addlinespace
\multirow{9}{*}{\shortstack[c]{Number of\\ Recovered Allocations}}    & 0.10 & 0.19   & 570 $\pm$ 82 & 994 $\pm$ 8 & 55 $\pm$ 33 & 32 $\pm$ 19 \\
 & 0.10 & 0.32    & 652 $\pm$ 80 & 997 $\pm$ 4 & 108 $\pm$ 54 & 18 $\pm$ 12 \\
 & 0.10 & 0.55    & 691 $\pm$ 52 & 1000 $\pm$ 0 & 99 $\pm$ 32 & 11 $\pm$ 6 \\
 & 0.25 & 0.19    & 392 $\pm$ 61 & 994 $\pm$ 8 & 49 $\pm$ 24 & 89 $\pm$ 18 \\
 & 0.25 & 0.32   & 502 $\pm$ 81 & 997 $\pm$ 4 & 75 $\pm$ 46 & 54 $\pm$ 21 \\
 & 0.25 & 0.55  & 734 $\pm$ 53 & 1000 $\pm$ 0 & 184 $\pm$ 50 & 26 $\pm$ 11 \\
 & 0.50 & 0.19    & 331 $\pm$ 63 & 995 $\pm$ 8 & 75 $\pm$ 36 & 84 $\pm$ 17 \\
 & 0.50 & 0.32    & 373 $\pm$ 68 & 997 $\pm$ 4 & 62 $\pm$ 41 & 68 $\pm$ 11 \\
 & 0.50 & 0.55   & 460 $\pm$ 69 & 1000 $\pm$ 0 & 74 $\pm$ 44 & 31 $\pm$ 12 \\
\addlinespace
\midrule
\addlinespace
\multirow{9}{*}{\shortstack[c]{Allocation Utility\\ ($k'/k$)}}   & 0.10 & 0.19    & 0.840 $\pm$ 0.031 & 0.851 $\pm$ 0.032 & 0.852 $\pm$ 0.034 & 0.852 $\pm$ 0.034 \\
 & 0.10 & 0.32    & 0.924 $\pm$ 0.022 & 0.930 $\pm$ 0.024 & 0.931 $\pm$ 0.025 & 0.931 $\pm$ 0.026 \\
 & 0.10 & 0.55    & 0.973 $\pm$ 0.011 & 0.975 $\pm$ 0.013 & 0.975 $\pm$ 0.015 & 0.974 $\pm$ 0.016 \\
 & 0.25 & 0.19    & 0.656 $\pm$ 0.039 & 0.658 $\pm$ 0.039 & 0.659 $\pm$ 0.039 & 0.659 $\pm$ 0.039 \\
 & 0.25 & 0.32    & 0.873 $\pm$ 0.020 & 0.877 $\pm$ 0.020 & 0.877 $\pm$ 0.021 & 0.877 $\pm$ 0.021 \\
 & 0.25 & 0.55    & 0.950 $\pm$ 0.011 & 0.953 $\pm$ 0.012 & 0.952 $\pm$ 0.013 & 0.952 $\pm$ 0.013 \\
 & 0.50 & 0.19    & 0.366 $\pm$ 0.023 & 0.367 $\pm$ 0.023 & 0.367 $\pm$ 0.023 & 0.367 $\pm$ 0.023 \\
 & 0.50 & 0.32    & 0.602 $\pm$ 0.026 & 0.604 $\pm$ 0.026 & 0.605 $\pm$ 0.026 & 0.604 $\pm$ 0.026 \\
 & 0.50 & 0.55 & 0.902 $\pm$ 0.013 & 0.904 $\pm$ 0.013 & 0.905 $\pm$ 0.013 & 0.905 $\pm$ 0.013 \\

\addlinespace
\bottomrule
\end{tabular} 
\label{tab:all_perf}
\end{table*}

\begin{table*}[b!]
\small
\centering
\caption{Evaluating Recovered Allocations From the Rashomon Set (c.f. Table~\ref{tab:results_main})}
\begin{tabular}{c|cc|c|cccc} 
\toprule
\multirow{2}{*}{Metric (Mean $\pm$ SD)} & \multirow{2}{*}{$k/n$} & \multirow{2}{*}{$n'/n$} & All Equal-Utility & Different & Bootstrapped & Shuffled & Weight \\
& & & Allocations & Feature Subsets & Training Data & Data Order & Perturbation \\
\toprule
\addlinespace
\multirow{9}{*}{\shortstack{Least\\Discriminatory\\Allocation $\dagger$}}  & 0.10 & 0.19 & 0.90 $\pm$ 0.08 & 0.97 $\pm$ 0.14 & 1.00 $\pm$ 0.12 & 1.09 $\pm$ 0.12 & 1.10 $\pm$ 0.13 \\
 & 0.10 & 0.32 & 0.95 $\pm$ 0.09 & 0.92 $\pm$ 0.14 & 0.99 $\pm$ 0.13 & 1.08 $\pm$ 0.12 & 1.11 $\pm$ 0.13 \\
 & 0.10 & 0.55 & 1.00 $\pm$ 0.09 & 0.90 $\pm$ 0.14 & 0.99 $\pm$ 0.15 & 1.09 $\pm$ 0.13 & 1.13 $\pm$ 0.13 \\
 & 0.25 & 0.19 & 1.06 $\pm$ 0.09 & 1.12 $\pm$ 0.13 & 1.14 $\pm$ 0.12 & 1.19 $\pm$ 0.13 & 1.16 $\pm$ 0.12 \\
 & 0.25 & 0.32 & 0.86 $\pm$ 0.08 & 1.12 $\pm$ 0.13 & 1.12 $\pm$ 0.12 & 1.19 $\pm$ 0.13 & 1.18 $\pm$ 0.12 \\
 & 0.25 & 0.55 & 0.92 $\pm$ 0.08 & 1.10 $\pm$ 0.14 & 1.12 $\pm$ 0.13 & 1.18 $\pm$ 0.13 & 1.20 $\pm$ 0.13 \\
 & 0.50 & 0.19 & 1.20 $\pm$ 0.12 & 1.25 $\pm$ 0.14 & 1.25 $\pm$ 0.13 & 1.30 $\pm$ 0.14 & 1.28 $\pm$ 0.14 \\
 & 0.50 & 0.32 & 1.21 $\pm$ 0.12 & 1.26 $\pm$ 0.14 & 1.25 $\pm$ 0.13 & 1.31 $\pm$ 0.14 & 1.28 $\pm$ 0.14 \\
 & 0.50 & 0.55 & 0.90 $\pm$ 0.10 & 1.25 $\pm$ 0.14 & 1.24 $\pm$ 0.13 & 1.30 $\pm$ 0.14 & 1.29 $\pm$ 0.14 \\

 \addlinespace
 \midrule
 \addlinespace
 \multirow{9}{*}{\shortstack{Outcome\\Homogenization $\ddagger$}}    & 0.10 & 0.19 & 0.88 $\pm$ 0.01 & 0.96 $\pm$ 0.00 & 0.98 $\pm$ 0.00 & 0.99 $\pm$ 0.00 & 0.99 $\pm$ 0.00 \\
 & 0.10 & 0.32 & 0.85 $\pm$ 0.00 & 0.95 $\pm$ 0.00 & 0.97 $\pm$ 0.00 & 0.99 $\pm$ 0.00 & 1.00 $\pm$ 0.00 \\
 & 0.10 & 0.55 & 0.83 $\pm$ 0.00 & 0.94 $\pm$ 0.00 & 0.96 $\pm$ 0.00 & 0.99 $\pm$ 0.00 & 1.00 $\pm$ 0.00 \\
 & 0.25 & 0.19 & 0.81 $\pm$ 0.01 & 0.96 $\pm$ 0.01 & 0.97 $\pm$ 0.00 & 0.99 $\pm$ 0.00 & 0.98 $\pm$ 0.00 \\
 & 0.25 & 0.32 & 0.80 $\pm$ 0.01 & 0.95 $\pm$ 0.00 & 0.97 $\pm$ 0.00 & 0.99 $\pm$ 0.00 & 0.99 $\pm$ 0.00 \\
 & 0.25 & 0.55 & 0.71 $\pm$ 0.01 & 0.92 $\pm$ 0.01 & 0.95 $\pm$ 0.00 & 0.98 $\pm$ 0.00 & 0.99 $\pm$ 0.00 \\
 & 0.50 & 0.19 & 0.60 $\pm$ 0.01 & 0.94 $\pm$ 0.01 & 0.95 $\pm$ 0.01 & 0.99 $\pm$ 0.00 & 0.97 $\pm$ 0.00 \\
 & 0.50 & 0.32 & 0.69 $\pm$ 0.01 & 0.95 $\pm$ 0.01 & 0.96 $\pm$ 0.01 & 0.99 $\pm$ 0.00 & 0.98 $\pm$ 0.00 \\
 & 0.50 & 0.55 & 0.76 $\pm$ 0.02 & 0.94 $\pm$ 0.01 & 0.96 $\pm$ 0.01 & 0.99 $\pm$ 0.00 & 0.99 $\pm$ 0.00 \\

 \addlinespace
 \midrule
 \addlinespace
 \multirow{9}{*}{\shortstack{Age\\Homogenization $\star$}}   & 0.10 & 0.19 & 2.46 $\pm$ 0.05 & 2.19 $\pm$ 0.10 & 2.27 $\pm$ 0.08 & 2.30 $\pm$ 0.09 & 2.30 $\pm$ 0.09 \\
 & 0.10 & 0.32 & 2.49 $\pm$ 0.04 & 2.04 $\pm$ 0.10 & 2.09 $\pm$ 0.09 & 2.24 $\pm$ 0.10 & 2.28 $\pm$ 0.10 \\
 & 0.10 & 0.55 & 2.56 $\pm$ 0.03 & 1.82 $\pm$ 0.11 & 1.84 $\pm$ 0.11 & 2.24 $\pm$ 0.09 & 2.28 $\pm$ 0.10 \\
 & 0.25 & 0.19 & 2.50 $\pm$ 0.04 & 2.43 $\pm$ 0.05 & 2.42 $\pm$ 0.05 & 2.39 $\pm$ 0.05 & 2.40 $\pm$ 0.05 \\
 & 0.25 & 0.32 & 2.49 $\pm$ 0.04 & 2.35 $\pm$ 0.06 & 2.33 $\pm$ 0.05 & 2.36 $\pm$ 0.05 & 2.36 $\pm$ 0.05 \\
 & 0.25 & 0.55 & 2.56 $\pm$ 0.03 & 2.20 $\pm$ 0.05 & 2.21 $\pm$ 0.06 & 2.31 $\pm$ 0.05 & 2.33 $\pm$ 0.06 \\
 & 0.50 & 0.19 & 2.55 $\pm$ 0.03 & 2.53 $\pm$ 0.03 & 2.48 $\pm$ 0.03 & 2.46 $\pm$ 0.04 & 2.47 $\pm$ 0.04 \\
 & 0.50 & 0.32 & 2.52 $\pm$ 0.03 & 2.51 $\pm$ 0.03 & 2.48 $\pm$ 0.03 & 2.47 $\pm$ 0.04 & 2.47 $\pm$ 0.04 \\
 & 0.50 & 0.55 & 2.55 $\pm$ 0.03 & 2.49 $\pm$ 0.03 & 2.46 $\pm$ 0.04 & 2.45 $\pm$ 0.04 & 2.45 $\pm$ 0.04 \\

\addlinespace
\bottomrule
\addlinespace
\multicolumn{8}{l}{$\dagger$ The least discriminatory allocation shows the lowest ratio of active chronic illnesses between selected Black \& White patients.} \\
\multicolumn{8}{l}{$\ddagger$ Outcome homogenization shows the probability of an individual receiving the same outcome under 2 random allocations.} \\
\multicolumn{8}{l}{$\star$ Age homogenization shows the Shannon entropy of age brackets across selected patients ($\uparrow$ entropy = $\downarrow$ homogenization).}

\end{tabular} 
\label{tab:all_results}
\end{table*}

\begin{table*}[b!]
\small
\centering
\caption{Outcome Homogenization Across the Rashomon Set (c.f. Figure~\ref{fig:results_oh})}
\begin{tabular}{c|cc|cccc} 
\toprule
Proportion of Qualified Individuals & \multirow{2}{*}{$k/n$} & \multirow{2}{*}{$n'/n$} & Different & Bootstrapped & Shuffled & Weight \\
(Mean $\pm$ SD) & &  & Feature Subsets & Training Data & Data Order & Perturbation \\
\toprule
\addlinespace
\multirow{9}{*}{\shortstack[c]{Rejected by all\\ Rashomon Models}}  & 0.10 & 0.19 & 0.182 $\pm$ 0.028 & 0.323 $\pm$ 0.034 & 0.509 $\pm$ 0.028 & 0.499 $\pm$ 0.035 \\
 & 0.10 & 0.32 & 0.313 $\pm$ 0.025 & 0.451 $\pm$ 0.028 & 0.645 $\pm$ 0.027 & 0.683 $\pm$ 0.024 \\
 & 0.10 & 0.55 & 0.462 $\pm$ 0.031 & 0.592 $\pm$ 0.022 & 0.775 $\pm$ 0.014 & 0.811 $\pm$ 0.007 \\
 & 0.25 & 0.19 & 0.050 $\pm$ 0.017 & 0.077 $\pm$ 0.018 & 0.119 $\pm$ 0.021 & 0.098 $\pm$ 0.020 \\
 & 0.25 & 0.32 & 0.103 $\pm$ 0.019 & 0.163 $\pm$ 0.021 & 0.285 $\pm$ 0.026 & 0.265 $\pm$ 0.029 \\
 & 0.25 & 0.55 & 0.217 $\pm$ 0.028 & 0.311 $\pm$ 0.020 & 0.506 $\pm$ 0.017 & 0.541 $\pm$ 0.015 \\
 & 0.50 & 0.19 & 0.010 $\pm$ 0.009 & 0.014 $\pm$ 0.009 & 0.025 $\pm$ 0.012 & 0.020 $\pm$ 0.011 \\
 & 0.50 & 0.32 & 0.018 $\pm$ 0.008 & 0.032 $\pm$ 0.010 & 0.046 $\pm$ 0.012 & 0.038 $\pm$ 0.011 \\
 & 0.50 & 0.55 & 0.044 $\pm$ 0.015 & 0.081 $\pm$ 0.011 & 0.154 $\pm$ 0.019 & 0.142 $\pm$ 0.023 \\

\addlinespace
\midrule
\addlinespace
\multirow{9}{*}{\shortstack[c]{Multiple Outcomes\\ From Rashomon Models}}   & 0.10 & 0.19 & 0.797 $\pm$ 0.038 & 0.521 $\pm$ 0.040 & 0.085 $\pm$ 0.031 & 0.081 $\pm$ 0.032 \\
 & 0.10 & 0.32 & 0.678 $\pm$ 0.026 & 0.489 $\pm$ 0.031 & 0.117 $\pm$ 0.035 & 0.041 $\pm$ 0.023 \\
 & 0.10 & 0.55 & 0.535 $\pm$ 0.032 & 0.376 $\pm$ 0.024 & 0.085 $\pm$ 0.023 & 0.018 $\pm$ 0.008 \\
 & 0.25 & 0.19 & 0.683 $\pm$ 0.194 & 0.133 $\pm$ 0.025 & 0.015 $\pm$ 0.009 & 0.053 $\pm$ 0.018 \\
 & 0.25 & 0.32 & 0.797 $\pm$ 0.064 & 0.408 $\pm$ 0.041 & 0.054 $\pm$ 0.018 & 0.083 $\pm$ 0.024 \\
 & 0.25 & 0.55 & 0.731 $\pm$ 0.056 & 0.497 $\pm$ 0.024 & 0.115 $\pm$ 0.020 & 0.040 $\pm$ 0.014 \\
 & 0.50 & 0.19 & 0.477 $\pm$ 0.144 & 0.057 $\pm$ 0.017 & 0.005 $\pm$ 0.005 & 0.015 $\pm$ 0.009 \\
 & 0.50 & 0.32 & 0.488 $\pm$ 0.140 & 0.075 $\pm$ 0.015 & 0.006 $\pm$ 0.005 & 0.023 $\pm$ 0.008 \\
 & 0.50 & 0.55 & 0.660 $\pm$ 0.079 & 0.307 $\pm$ 0.035 & 0.032 $\pm$ 0.011 & 0.048 $\pm$ 0.020 \\

\addlinespace
\midrule
\addlinespace
\multirow{9}{*}{\shortstack[c]{Accepted by all\\ Rashomon Models}}      & 0.10 & 0.19 & 0.021 $\pm$ 0.022 & 0.157 $\pm$ 0.038 & 0.406 $\pm$ 0.038 & 0.419 $\pm$ 0.031 \\
 & 0.10 & 0.32 & 0.010 $\pm$ 0.007 & 0.061 $\pm$ 0.018 & 0.239 $\pm$ 0.023 & 0.277 $\pm$ 0.015 \\
 & 0.10 & 0.55 & 0.003 $\pm$ 0.003 & 0.031 $\pm$ 0.008 & 0.140 $\pm$ 0.012 & 0.171 $\pm$ 0.006 \\
 & 0.25 & 0.19 & 0.267 $\pm$ 0.189 & 0.790 $\pm$ 0.028 & 0.866 $\pm$ 0.022 & 0.849 $\pm$ 0.024 \\
 & 0.25 & 0.32 & 0.101 $\pm$ 0.053 & 0.429 $\pm$ 0.044 & 0.661 $\pm$ 0.033 & 0.651 $\pm$ 0.028 \\
 & 0.25 & 0.55 & 0.052 $\pm$ 0.033 & 0.192 $\pm$ 0.016 & 0.379 $\pm$ 0.015 & 0.419 $\pm$ 0.013 \\
 & 0.50 & 0.19 & 0.512 $\pm$ 0.140 & 0.929 $\pm$ 0.018 & 0.971 $\pm$ 0.013 & 0.966 $\pm$ 0.014 \\
 & 0.50 & 0.32 & 0.494 $\pm$ 0.136 & 0.894 $\pm$ 0.017 & 0.948 $\pm$ 0.013 & 0.939 $\pm$ 0.013 \\
 & 0.50 & 0.55 & 0.296 $\pm$ 0.067 & 0.613 $\pm$ 0.033 & 0.815 $\pm$ 0.019 & 0.810 $\pm$ 0.017 \\

\addlinespace
\bottomrule
\end{tabular} 
\label{tab:outcome_homog}
\end{table*}

\begin{table*}[b!]
\small
\centering
\caption{Evaluating Stochastic Mappings Between Predictions \& Decisions}
\begin{tabular}{c|cc|cc|cccc} 
\toprule
\multirow{2}{*}{Metric (Mean $\pm$ SD)} & \multirow{2}{*}{$k/n$} & \multirow{2}{*}{$n'/n$} & All Equal-Utility & Avg. Recovered & \multicolumn{2}{c}{Decision-Boundary} & \multicolumn{2}{c}{Sigmoid-Logit} \\
& & & Allocations & Rashomon Set & (0.25$k$, 0.50$k$) & (0.50$k$, 1.0$k$) & ($v=$2) & ($v=$5) \\
\toprule
\addlinespace
\multirow{9}{*}{\shortstack{Allocation \\ Utility (($k'/k$)}}  
  & 0.10 & 0.19 & -  & 0.849 $\pm$ 0.033 & 0.850 $\pm$ 0.031 & 0.839 $\pm$ 0.030 & 0.832 $\pm$ 0.028 & 0.851 $\pm$ 0.032 \\
 & 0.10 & 0.32 & -  & 0.929 $\pm$ 0.024 & 0.929 $\pm$ 0.026 & 0.924 $\pm$ 0.023 & 0.893 $\pm$ 0.019 & 0.917 $\pm$ 0.022 \\
 & 0.10 & 0.55 & -  & 0.974 $\pm$ 0.014 & 0.976 $\pm$ 0.013 & 0.974 $\pm$ 0.012 & 0.926 $\pm$ 0.011 & 0.940 $\pm$ 0.012 \\
 & 0.25 & 0.19 & -  & 0.658 $\pm$ 0.039 & 0.649 $\pm$ 0.034 & 0.623 $\pm$ 0.031 & 0.653 $\pm$ 0.038 & 0.658 $\pm$ 0.039 \\
 & 0.25 & 0.32 & -  & 0.876 $\pm$ 0.021 & 0.869 $\pm$ 0.020 & 0.834 $\pm$ 0.021 & 0.851 $\pm$ 0.019 & 0.866 $\pm$ 0.019 \\
 & 0.25 & 0.55 & -  & 0.952 $\pm$ 0.012 & 0.952 $\pm$ 0.012 & 0.943 $\pm$ 0.011 & 0.898 $\pm$ 0.012 & 0.898 $\pm$ 0.013 \\
 & 0.50 & 0.19 & -  & 0.367 $\pm$ 0.023 & 0.365 $\pm$ 0.023 & 0.363 $\pm$ 0.023 & 0.364 $\pm$ 0.023 & 0.367 $\pm$ 0.023 \\
 & 0.50 & 0.32 & -  & 0.604 $\pm$ 0.026 & 0.599 $\pm$ 0.025 & 0.593 $\pm$ 0.025 & 0.601 $\pm$ 0.025 & 0.603 $\pm$ 0.026 \\
 & 0.50 & 0.55 & -  & 0.904 $\pm$ 0.013 & 0.885 $\pm$ 0.013 & 0.850 $\pm$ 0.013 & 0.868 $\pm$ 0.013 & 0.888 $\pm$ 0.013 \\

\addlinespace
\midrule
\addlinespace
\multirow{9}{*}{\shortstack{Least\\Discriminatory\\Allocation $\dagger$}}   & 0.10 & 0.19 & 0.90 $\pm$ 0.08 & 1.04 $\pm$ 0.13 & 1.04 $\pm$ 0.12 & 1.00 $\pm$ 0.13 & 0.94 $\pm$ 0.12 & 1.02 $\pm$ 0.12 \\
 & 0.10 & 0.32 & 0.95 $\pm$ 0.09 & 1.03 $\pm$ 0.13 & 1.07 $\pm$ 0.13 & 1.02 $\pm$ 0.14 & 0.78 $\pm$ 0.12 & 0.87 $\pm$ 0.12 \\
 & 0.10 & 0.55 & 1.00 $\pm$ 0.09 & 1.03 $\pm$ 0.14 & 1.09 $\pm$ 0.16 & 1.02 $\pm$ 0.16 & 0.61 $\pm$ 0.12 & 0.66 $\pm$ 0.11 \\
 & 0.25 & 0.19 & 1.06 $\pm$ 0.09 & 1.15 $\pm$ 0.12 & 1.10 $\pm$ 0.11 & 1.08 $\pm$ 0.11 & 1.09 $\pm$ 0.11 & 1.13 $\pm$ 0.12 \\
 & 0.25 & 0.32 & 0.86 $\pm$ 0.08 & 1.15 $\pm$ 0.12 & 1.11 $\pm$ 0.11 & 1.09 $\pm$ 0.12 & 1.02 $\pm$ 0.11 & 1.04 $\pm$ 0.11 \\
 & 0.25 & 0.55 & 0.92 $\pm$ 0.08 & 1.15 $\pm$ 0.13 & 1.14 $\pm$ 0.13 & 1.11 $\pm$ 0.13 & 0.87 $\pm$ 0.11 & 0.87 $\pm$ 0.12 \\
 & 0.50 & 0.19 & 1.20 $\pm$ 0.12 & 1.27 $\pm$ 0.14 & 1.24 $\pm$ 0.13 & 1.23 $\pm$ 0.13 & 1.22 $\pm$ 0.13 & 1.25 $\pm$ 0.14 \\
 & 0.50 & 0.32 & 1.21 $\pm$ 0.12 & 1.27 $\pm$ 0.14 & 1.23 $\pm$ 0.13 & 1.22 $\pm$ 0.13 & 1.23 $\pm$ 0.13 & 1.25 $\pm$ 0.13 \\
 & 0.50 & 0.55 & 0.90 $\pm$ 0.10 & 1.27 $\pm$ 0.14 & 1.23 $\pm$ 0.13 & 1.23 $\pm$ 0.14 & 1.13 $\pm$ 0.13 & 1.14 $\pm$ 0.13 \\

\addlinespace
\midrule
\addlinespace
\multirow{9}{*}{\shortstack{Outcome\\Homogenization $\ddagger$}} & 0.10 & 0.19 & 0.88 $\pm$ 0.01 & 0.98 $\pm$ 0.00 & 0.97 $\pm$ 0.00 & 0.95 $\pm$ 0.00 & 0.94 $\pm$ 0.01 & 0.97 $\pm$ 0.00 \\
 & 0.10 & 0.32 & 0.85 $\pm$ 0.00 & 0.98 $\pm$ 0.00 & 0.97 $\pm$ 0.00 & 0.95 $\pm$ 0.00 & 0.89 $\pm$ 0.00 & 0.93 $\pm$ 0.01 \\
 & 0.10 & 0.55 & 0.83 $\pm$ 0.00 & 0.97 $\pm$ 0.00 & 0.97 $\pm$ 0.00 & 0.95 $\pm$ 0.00 & 0.85 $\pm$ 0.00 & 0.86 $\pm$ 0.00 \\
 & 0.25 & 0.19 & 0.81 $\pm$ 0.01 & 0.98 $\pm$ 0.00 & 0.94 $\pm$ 0.00 & 0.90 $\pm$ 0.00 & 0.94 $\pm$ 0.01 & 0.97 $\pm$ 0.00 \\
 & 0.25 & 0.32 & 0.80 $\pm$ 0.01 & 0.98 $\pm$ 0.00 & 0.94 $\pm$ 0.00 & 0.88 $\pm$ 0.00 & 0.89 $\pm$ 0.01 & 0.93 $\pm$ 0.02 \\
 & 0.25 & 0.55 & 0.71 $\pm$ 0.01 & 0.96 $\pm$ 0.00 & 0.94 $\pm$ 0.00 & 0.88 $\pm$ 0.00 & 0.73 $\pm$ 0.01 & 0.73 $\pm$ 0.01 \\
 & 0.50 & 0.19 & 0.60 $\pm$ 0.01 & 0.96 $\pm$ 0.01 & 0.89 $\pm$ 0.00 & 0.84 $\pm$ 0.01 & 0.85 $\pm$ 0.01 & 0.95 $\pm$ 0.01 \\
 & 0.50 & 0.32 & 0.69 $\pm$ 0.01 & 0.97 $\pm$ 0.01 & 0.89 $\pm$ 0.00 & 0.84 $\pm$ 0.01 & 0.90 $\pm$ 0.01 & 0.95 $\pm$ 0.01 \\
 & 0.50 & 0.55 & 0.76 $\pm$ 0.02 & 0.97 $\pm$ 0.01 & 0.89 $\pm$ 0.00 & 0.81 $\pm$ 0.01 & 0.84 $\pm$ 0.02 & 0.89 $\pm$ 0.02 \\

\addlinespace
\midrule
\addlinespace

\multirow{9}{*}{\shortstack{Age\\Homogenization $\star$}}
 & 0.10 & 0.19 & 2.46 $\pm$ 0.05 & 2.26 $\pm$ 0.09 & 2.28 $\pm$ 0.09 & 2.30 $\pm$ 0.08 & 2.32 $\pm$ 0.07 & 2.27 $\pm$ 0.09 \\
 & 0.10 & 0.32 & 2.49 $\pm$ 0.04 & 2.16 $\pm$ 0.10 & 2.10 $\pm$ 0.11 & 2.14 $\pm$ 0.09 & 2.27 $\pm$ 0.06 & 2.17 $\pm$ 0.08 \\
 & 0.10 & 0.55 & 2.56 $\pm$ 0.03 & 2.04 $\pm$ 0.10 & 1.83 $\pm$ 0.13 & 1.92 $\pm$ 0.10 & 2.36 $\pm$ 0.03 & 2.27 $\pm$ 0.05 \\
 & 0.25 & 0.19 & 2.50 $\pm$ 0.04 & 2.41 $\pm$ 0.05 & 2.43 $\pm$ 0.05 & 2.44 $\pm$ 0.04 & 2.44 $\pm$ 0.05 & 2.43 $\pm$ 0.05 \\
 & 0.25 & 0.32 & 2.49 $\pm$ 0.04 & 2.35 $\pm$ 0.05 & 2.38 $\pm$ 0.05 & 2.43 $\pm$ 0.04 & 2.42 $\pm$ 0.04 & 2.37 $\pm$ 0.05 \\
 & 0.25 & 0.55 & 2.56 $\pm$ 0.03 & 2.26 $\pm$ 0.06 & 2.23 $\pm$ 0.06 & 2.27 $\pm$ 0.05 & 2.50 $\pm$ 0.03 & 2.49 $\pm$ 0.03 \\
 & 0.50 & 0.19 & 2.55 $\pm$ 0.03 & 2.48 $\pm$ 0.04 & 2.51 $\pm$ 0.03 & 2.51 $\pm$ 0.04 & 2.53 $\pm$ 0.03 & 2.48 $\pm$ 0.04 \\
 & 0.50 & 0.32 & 2.52 $\pm$ 0.03 & 2.48 $\pm$ 0.04 & 2.52 $\pm$ 0.03 & 2.52 $\pm$ 0.03 & 2.52 $\pm$ 0.03 & 2.50 $\pm$ 0.04 \\
 & 0.50 & 0.55 & 2.55 $\pm$ 0.03 & 2.46 $\pm$ 0.04 & 2.50 $\pm$ 0.03 & 2.52 $\pm$ 0.03 & 2.56 $\pm$ 0.03 & 2.56 $\pm$ 0.03 \\

\addlinespace
\bottomrule
\addlinespace
\multicolumn{9}{l}{$\dagger$ The least discriminatory allocation shows the lowest ratio of active chronic illnesses between selected Black \& White patients.} \\
\multicolumn{9}{l}{$\ddagger$ Outcome homogenization shows the probability of an individual receiving the same outcome under 2 random allocations.} \\
\multicolumn{9}{l}{$\star$ Age homogenization shows the Shannon entropy of age brackets across selected patients ($\uparrow$ entropy = $\downarrow$ homogenization).}

\end{tabular} 
\label{tab:all_random_results}
\end{table*}

\clearpage

\end{document}